\documentclass[10pt]{article}

\usepackage{graphicx}
\usepackage{amsmath,amssymb}
\usepackage[margin=0.75in]{geometry}
\usepackage[dvipsnames,table]{xcolor}
\usepackage{booktabs}
\usepackage{nicematrix}
\usepackage{caption}
\usepackage{subcaption}
\usepackage{microtype}
\usepackage{url}
\usepackage[numbers,sort&compress]{natbib}
\usepackage[colorlinks=true,linkcolor=blue,citecolor=blue,urlcolor=blue,breaklinks=true]{hyperref}

\graphicspath{{./}}
\renewcommand{\arraystretch}{1.2}

\newcommand{\startfullwidthfigure}{\clearpage\onecolumn}
\newcommand{\finishfullwidthfigure}{\clearpage\twocolumn}

\begin{document}
\raggedbottom
\vbadness=10000
\hbadness=10000

\begin{titlepage}
\begin{center}
{\Large\bfseries Scaling laws in complex component systems as consequences of heterogeneous sampling\par}

\vspace{1.0\baselineskip}

Luca Allegri$^{1,2,*\dagger}$, Johannes Nauta$^{1,*\dagger}$, and Manlio De Domenico$^{1,2,*}$

\vspace{0.8\baselineskip}

\begin{minipage}{0.98\textwidth}
\centering
\small
$^{1}$Department of Physics and Astronomy ``Galileo Galilei'', University of Padua,\par
via F. Marzolo 8, 35131 Padova, Italy\par
$^{2}$INFN, Sez. di Padova, Italy\par
\vspace{0.75\baselineskip}
$^{*}$Correspondence: luca.allegri.1@phd.unipd.it; johannes.nauta@unipd.it; manlio.dedomenico@unipd.it\par
$^{\dagger}$These authors contributed equally to this work.
\end{minipage}
\end{center}

\vspace{1.0\baselineskip}

\noindent\textbf{Preprint note.}
This manuscript is the original version submitted to Nature Communications and has not yet undergone peer review.

\vspace{1.0\baselineskip}

\begin{abstract}
Complex component systems are collections of discrete units such as species, words, genes, whose observed realizations are naturally summarized by component counts.
Many empirical laws have been observed in those systems, such as Taylor's law, Zipf's law, and Heaps' law, and domain-specific mechanisms are often employed to explain their emergence but, despite their ubiquity, a unifying framework remains elusive.
In this work, we propose a null model showing that, under heterogeneous latent rates and finite sampling, several commonly observed scaling relations can arise without invoking domain-specific mechanisms.

Taylor's law, for instance, reflects a crossover between sampling noise and genuine system heterogeneity and it is largely insensitive to the detailed latent distribution, while Zipf's and Heaps' laws arise from the convergence of order statistics and distinct component counts under heavy-tailed but otherwise generic priors.

Our work thus suggests that these ubiquitous patterns are better interpreted as a transient sign of statistical convergence instead of fundamental principles that require tailored generative explanations.
\end{abstract}


\end{titlepage}

\setcounter{page}{2}
\setcounter{footnote}{0}
\twocolumn
\bigskip

Complex systems display a remarkable level of statistical regularity~\cite{brown2000scaling,altmann2024statistical}.
These empirical regularities have been investigated under the guise of ``laws'', yet their origins and implications are still debated.
For instance, in 1961, Taylor~\cite{taylor1961aggregation} described a power law relation between fluctuations in the number of individuals of a species and their mean.
Whereas this pattern, known as \emph{Taylor's law}~\cite{southwood1988ecological}, was initially rooted in ecology~\cite{marquet2005scaling}, it has since been observed in a diverse set of complex systems in biology~\cite{west1999origin}, natural phenomena~\cite{dahlstedt2005fluctuation,tippett2016tornado}, physics~\cite{vaughan2008studying}, finance~\cite{eisler2008fluctuation}, linguistics~\cite{gerlach2014scaling,tanaka-ishii2018taylors} and human activities~\cite{tria2020taylors}, among others.
Since its original appearance, specific models have been put forward to explain Taylor's law in ecological systems~\cite{giometto2015sample,grilli2020macroecological} and beyond~\cite{james2018zipfs,eisler2008fluctuation}.
However, its ubiquity across distinct domains suggests a more general statistical origin.
Similarly, other laws related to order statistics and vocabulary growth, known as Zipf's and Heaps' law respectively~\cite{zipf1942unity,herdan1958relation,heaps1978information,gabaix1999zipfs,newman2005power,fontclos2013scaling,james2018zipfs,mazzolini2018heaps,mazzarisi2021maximal}, exhibit similar omnipresence and are often observed simultaneously.

Their recurrent and simultaneous appearance raises the question of whether a single set of minimal statistical assumptions can jointly generate them.

\begin{figure}[!ht]
  \centering
  \includegraphics[width=\textwidth,height=.58\textheight,keepaspectratio]{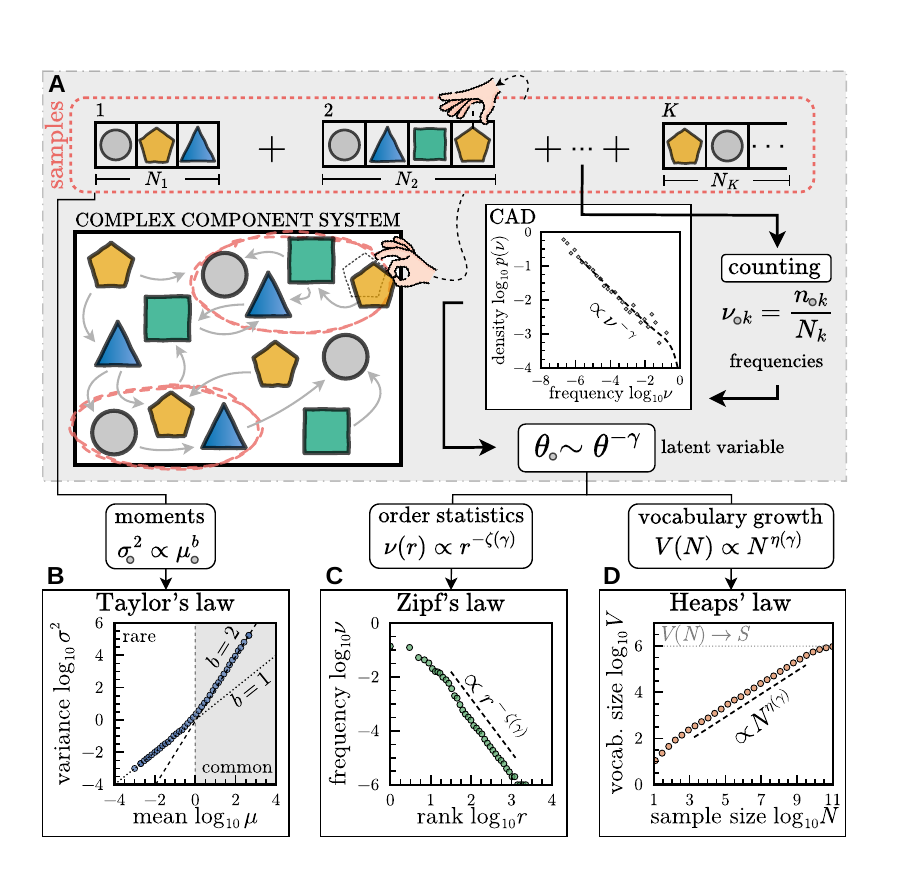}
  \caption{\textbf{Emergent universality from statistical convergence.}
  Taylor's law, Zipf's law, and Heaps' law emerge from sampling large component systems with latent variables $\theta \sim \theta^{-\gamma}$.
  (\textbf{A})~Taylor's law states that (sample) mean and variance scale as $\sigma^2 \propto \mu^b$.
  For rare components with $\mu_i \ll 1$, sampling noise dominates and we predict $b \rightarrow 1$.
  Common components with $\mu_i \gg 1$ inherit the fluctuations of the latent variables, leading to $b \rightarrow 2$.
  (\textbf{B})~Zipf's law states that frequencies (relative abundances) scale with rank as $\nu \propto r^{-\zeta}$, where $\zeta = 1/(\gamma-1)$.
  Low rank (common) components homogenize as their distribution across concatenated observations becomes more uniform, leading to a plateau for low ranks.
  (\textbf{C})~Heaps' law states that vocabulary size and observation length scale as $V(N) \propto N^{\eta}$.
  For $N \ll \varphi$ we predict linear scaling, $V \propto N$, whereas Heaps' law applies in the intermediate regime $1 \ll N \ll \varphi$.
  Eventually, $V$ must saturate due to the inherent finite number of components $S$, however note that this saturation is often not observed in empirical settings.}
  \label{fig:illustration}
\end{figure}

What connects these laws is that they arise from observations of complex \emph{component systems}~\cite{mazzolini2018statistics}.
Such systems are described by counts of their constituent units, such as species in an ecosystem, words in a text, genes in a genome, or even LEGO pieces in a set.
In practice, such systems are exclusively accessed through coarse-grained observations in the form of component counts (Fig.~\ref{fig:illustration}).
Perhaps because there is little discourse on the nature of observations of component systems, explanations of the laws that emerge from them tend to focus on bespoke generative models, such as stochastic growth models~\cite{miller1957effects,eliazar2011growth,giometto2015sample,mazzolini2018heaps}, preferential attachment schemes~\cite{yule1925iia,simon1955class,yamasaki2006preferential}, or evolutionary~\cite{demarzo2021dynamical} or innovation processes~\cite{tria2018zipfs}.
While these approaches differ in their assumptions, they share a focus on mechanism-level explanations often foregoing possible effects of sampling (but see~\cite{eisler2008fluctuation} for an exception).
As a result this raises a complementary question: which observed regularities reflect intrinsic properties of the underlying system, and which arise generically from the simple act of sampling heterogeneous systems?

Here, we address this question by considering a minimal probabilistic framework for sampling from heterogeneous component systems.
We show that, under broad and empirically testable conditions, namely heterogeneity in underlying frequencies and finite sampling, several canonical scaling laws arise jointly as different statistical summaries of a common latent structure~(Fig.~\ref{fig:illustration}).
In particular, Taylor's law follows \emph{locally} from a mean--variance decomposition in which each component has its own crossover between sampling noise and latent heterogeneity.
Aggregating components with different crossover scales then produces the apparent global power-law behavior usually reported in empirical Taylor-law analyses.
The same latent structure also constrains the abundance distributions observed within individual samples.
In particular, across diverse datasets spanning biology, language and finance, among others~(see~Supplementary Information 1), we find that component abundance distributions are heavy-tailed and that, within each dataset, their tail exponents fluctuate around a dataset-specific value.
This supports the interpretation that samples are finite observations of a shared latent abundance structure.
Once this structure is heavy-tailed, Zipf's and Heaps' laws follow from the known behavior of order statistics and distinct component counts under finite sampling.
These results can be formalized by noting that unordered component-count observations can be approximated as finite exchangeable sequences, leading naturally to latent-variable representations.

Altogether, our work reframes the widespread occurrence of scaling laws as a generic statistical consequence of sampling heterogeneous systems, providing a baseline against which genuinely informative system-specific properties can be identified.
At the same time, our framework does not explain why particular systems exhibit the observed degree or form of heterogeneity in component frequencies.
The origin of this heterogeneity, and in particular the values of specific distributional parameters such as tail exponents, may still depend on system-specific mechanisms.

\startfullwidthfigure
\finishfullwidthfigure

\section*{Results}
\label{sec:results}
\subsection*{Sampling systems with latent heterogeneity}\label{sec:latent}
To show how bespoke scaling laws naturally emerge from sampling heterogeneous systems, let us introduce a simple generic sampling scheme.
We consider an observer who repeatedly draws components from a system (Fig.~\ref{fig:illustration}).
Each observation or sample, which consists of $N$ independent draws, represents a different coarse-grained realization of the system and contains counts of components of each type (e.g., a species or a word).
The counts of type $i$ in sample $k$ are denoted with $n_{ik}$ and we assume that these counts depend on two factors, (i)~the \emph{sampling effort} $N_k$, representing the total number of draws in sample $k$ and, (ii)~an underlying \emph{latent intensity} $\theta_{ik}$, representing the true relative abundance of components of type $i$ in sample $k$.
We model the observed counts as Poisson random variables, $n_{ik} \sim \operatorname{Poisson}(\lambda_{ik})$ with latent rates $\lambda_{ik} = N_k \theta_{ik}$ (this is actually an approximation of a binomial process with parameters $N_k$ and $\theta_{ik} \ll 1$, see~Supplementary Information 4.6).
As these rates are unobservable, an observer instead probes the system using the \emph{relative counts} $\nu_{ik} = n_{ik} / N_k$ as coarse-grained representations with the aim of uncovering the \emph{latent distribution} $p(\theta)$ that captures system-specific structure.
Assuming fixed sample sizes for simplicity, we have $\operatorname{E}\!\left[\nu_i\right] = \operatorname{E}\!\left[\theta_i\right]$ and $\operatorname{Var}\!\left[\nu_i\right] = \operatorname{Var}\!\left[\theta_i\right] + \operatorname{E}\!\left[\theta_i\right] / N$.
For large sample sizes $N \to \infty$ and $\operatorname{E}\!\left[\theta_i\right] < \infty$, the variance of the relative counts converges to the variance of the latent frequencies, therefore, the distribution of observable relative abundances approaches that of the underlying latent variates.

Note that, by construction, our framework assumes component counts to be independent.
At first sight, this may appear to be a strong constraint.
The subtle point is that this independence is imposed at the level of the observational process, not necessarily at the level of the underlying system.
Components may still interact through microscopic dynamics, but the observer does not access these interactions directly: they only record whether a given component appears in a sampling event and then count the total number of appearances.

Thus, at the level of unordered component counts, the observation process is approximated by an exchangeable/conditionally independent sampling scheme.
As a result, microscopic correlations that are not resolved by the observer are represented effectively through the latent parameters generating the samples, and therefore remain hidden at this level of aggregation (see~Supplementary Information 2 for a detailed discussion).

\subsection*{Variability in latent variates and Taylor's law}\label{sec:taylor}
\begin{figure}[!ht]
  \centering
  \includegraphics[width=\textwidth,height=.50\textheight,keepaspectratio]{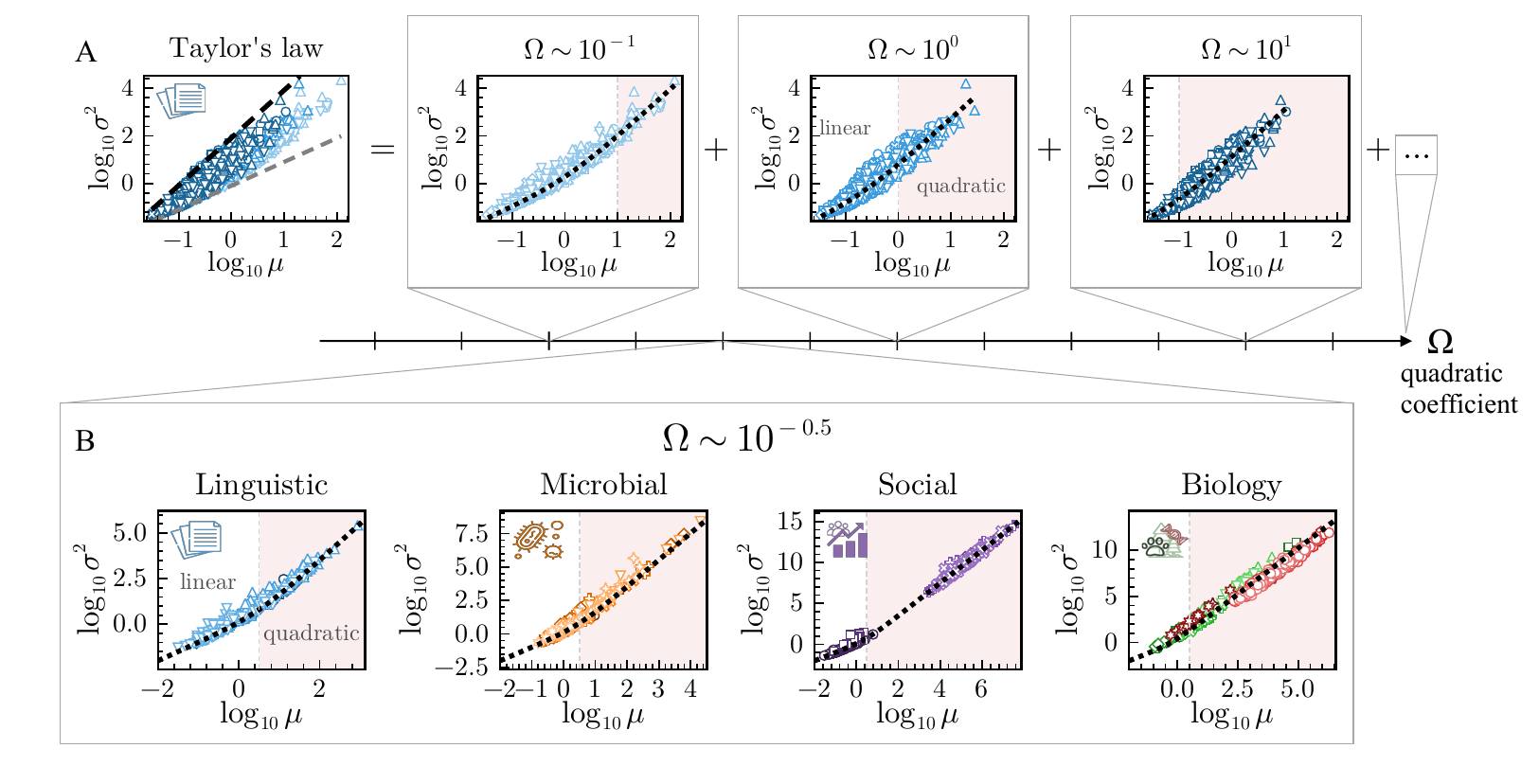}
  \caption{\textbf{Taylor's law as an aggregation of quadratic mean--variance relations.}
  (\textbf{A}) Taylor's law emerges as the result of aggregating component-wise quadratic relations of the form given in Eq.~\ref{eq:taylor}.
  Each component $i$ is associated with a quadratic coefficient $\Omega_i = \operatorname{Var}\!\left[\nu_i\right] / \operatorname{E}\!\left[\nu_i\right]^2$, where $\nu_i$ denotes the relative abundance of component $i$ across samples.
  When components with similar values of $\Omega_i$ are grouped together, the underlying quadratic scaling between the mean $\mu$ and the variance $\sigma^2$ becomes visible.
  The three central panels show linguistic data grouped by three representative values of the quadratic coefficient, $\Omega \sim 10^{-1}$, $\Omega \sim 10^0$, and $\Omega \sim 10^1$.
  In each group, the empirical points closely follow the theoretical curve $\sigma^2 = \mu + \Omega \mu^2$, computed using the corresponding value of $\Omega$ (black dotted line).
  The grey vertical dashed line marks the predicted crossover between the linear regime, $\sigma^2 \simeq \mu$, and the quadratic regime, $\sigma^2 \simeq \Omega \mu^2$.
  This crossover occurs at $\mu = \Omega^{-1}$, where the two contributions become equal.
  The scatter plot in the left represents the aggregate Taylor plot obtained by pooling together the same components shown in the panels on the right.
  In this sense, it is the superposition of many component groups, each following its own quadratic mean--variance relation with a different coefficient $\Omega$.
  The apparent Taylor's law observed in the aggregate therefore does not correspond to a single fundamental power law.
  Rather, it emerges from the sum of several local quadratic relations.
  Because different values of $\Omega$ shift the crossover scale $\mu=\Omega^{-1}$, their superposition produces an effective scaling with slope between the linear regime and the quadratic regime illustrated by the grey and black dashed reference lines.
  (\textbf{B}) The same component-wise quadratic scaling can be tested across different empirical domains by repeating the decomposition shown in \textbf{A}.
  Here, for each domain, components are grouped by their empirical quadratic coefficient $\Omega_i$, and one common coefficient bin, $\Omega \sim 10^{-0.5}$, is shown for comparison.
  The panels include linguistic systems in blue, microbial systems in orange, social systems in purple, and biological systems, including genetic data in red and ecological data in green.
  For all domains, the empirical points in the selected $\Omega$-bin follow the same theoretical curve with $\Omega \sim 10^{-0.5}$.
  The recurrence of this form across such different systems supports the interpretation that the scaling is not domain-specific.
  Rather, it reflects a common observational procedure: heterogeneous component abundances are sampled through a Poissonian counting process, producing the same mean--variance structure.
  A complete description of the data is found in~Supplementary Information 1.}
  \label{fig:taylor}
\end{figure}

Taylor's law relates mean and variance of component abundances across observations by a power law.
That is, if we let $\operatorname{E}\!\left[\cdot\right] = \operatorname{E}_k\!\left[\cdot\right]$ and $\operatorname{Var}\!\left[\cdot\right] = \operatorname{Var}_k\!\left[\cdot\right]$ denote expectation and variance over samples $k=1, 2, \ldots, K$, then Taylor's law reads 
\begin{equation}
  \label{eq:TL}
  \operatorname{Var}\!\left[n_i\right] \propto \operatorname{E}\!\left[n_i\right]^b,
\end{equation}
where $b$ is the scaling exponent, which is common for all components.
In the context of the sampling scheme introduced above, assuming a constant sampling effort $N_k = N$, the law of total variance directly yields the decomposition
\begin{equation}
  \label{eq:taylor}
  \operatorname{Var}\!\left[n_i\right] = \Omega[\theta_i]\operatorname{E}\!\left[n_i\right]^2 + \operatorname{E}\!\left[n_i\right],
\end{equation}
where $\Omega[\theta_i] = \operatorname{Var}\!\left[\theta_i\right] / \operatorname{E}\!\left[\theta_i\right]^2$ is the quadratic coefficient (see~Supplementary Information 4 for the full derivation and the generalization to the case of different sample sizes).
This expression separates the two contributions to the observed variability in component counts: the latter reflects intrinsic sampling noise, whereas the former reflects genuine heterogeneity in the latent intensities $\theta_{ik}$.
Consequently, their interplay naturally gives rise to two asymptotic regimes (Figs.~\ref{fig:illustration} and~\ref{fig:taylor}).
In particular, when the latent variability is negligible, $\Omega[\theta_i] \approx 0$, the variance is dominated by sampling noise and we have $\operatorname{Var}\!\left[n_i\right] \approx \operatorname{E}\!\left[n_i\right]$, which corresponds to Taylor's law with exponent $b=1$.
On the other hand, when the latent variability is large, $\Omega[\theta_i] \gg 1$, the variance is dominated by the variability in the latent intensities and we have $\operatorname{Var}\!\left[n_i\right] \approx \Omega[\theta_i]\operatorname{E}\!\left[n_i\right]^2$, which corresponds to Taylor's law with exponent $b=2$.

It is important to realize that Eq.~\ref{eq:taylor} is a \emph{local} relation: it holds for each component $i$ separately.
The local behaviour was tested by dividing samples into two independent groups multiple times and fitting the quadratic form in Eq.~\ref{eq:taylor} (see~Supplementary Information 4.1 for details).
From our analysis we found a relation $\Omega_{i,A} \approx \Omega_{i,B}$, where $A$ and $B$ refer to the two independent groups, for all datasets analysed, except for the GTEx data as discussed below.

The component-specific coefficient $\Omega[\theta_i]$ sets the crossover scale of the mean--variance relation.
Below this scale, the sampling term dominates, $\operatorname{Var}\!\left[n_i\right] \simeq \operatorname{E}\!\left[n_i\right]$; above it, the heterogeneity term dominates, $\operatorname{Var}\!\left[n_i\right] \simeq \Omega[\theta_i]\operatorname{E}\!\left[n_i\right]^2$.
This crossover occurs when the two contributions are comparable, namely at $\operatorname{E}\!\left[n_i\right] \simeq \Omega[\theta_i]^{-1}$.
Therefore, the quadratic form becomes visible only after conditioning on components with similar values of $\Omega[\theta_i]$ (Fig.~\ref{fig:taylor}).

In standard Taylor-law analyses, however, all components are usually shown together in a single log--log mean--variance plot~\cite{taylor1961aggregation,grilli2020macroecological,eisler2008fluctuation}.
This procedure aggregates components with different quadratic coefficients, and therefore with different crossover scales.
The resulting curve is not a single quadratic relation, but a superposition of local quadratic forms.
This superposition produces an apparent power law with an effective exponent $b$ between the linear regime, $b=1$, and the quadratic regime, $b=2$ (Fig.~\ref{fig:taylor}).
Remarkably, this is precisely the range of exponents commonly observed in empirical Taylor-law analyses across complex systems~\cite{eisler2008fluctuation}.
This interpretation also clarifies the role of rare components, namely components that are present in only a small fraction of samples.
In fact, it is known that including rare components tends to lower the measured Taylor exponent $b$ (see \cite{eisler2008fluctuation} and~Supplementary Information 4.2), because their empirical moments are strongly affected by the many zero counts entering their estimation, reflecting the fact that they are not found in many samples, where a zero count corresponds to a sample in which the component is not observed.
In our framework, this effect has a direct explanation: rare components typically lie closer to the sampling-dominated regime, where the linear contribution is dominant.
As a result, their inclusion gives more weight to components with an effective scaling closer to $b=1$, thus lowering the exponent measured from the aggregate Taylor plot.

Moreover, since our framework relies on a Poissonian sampling scheme, when latent variability remains nonzero and eventually dominates sampling noise, increasing sample size pushes the observed relation toward the quadratic regime.
This means that, in such conditions, Taylor's law is a \emph{transient} law, as increasing sample size pushes the mean-variance relation towards the quadratic regime, until the effective Taylor's exponent eventually becomes equal to $2$.
The effect of rare species and the dependence on sample size are reported in~Supplementary Information 4.2.

Noticeably, within our sampling model, the mean--variance decomposition does not depend on the detailed form of $p(\theta)$.
Thus, Taylor-like scaling can arise as a statistical baseline without requiring a domain-specific mechanism.

Rather, it is an effect created from sampling components with a latent structure and this provides a possible explanation of its ubiquity across the sciences (Fig.~\ref{fig:taylor}).

\startfullwidthfigure
\finishfullwidthfigure

\subsection*{Heavy-tailed latent distributions and Zipf's law}\label{sec:powerlaw}
The sampling framework introduced so far assumes that the latent intensities $\lambda_{ik}$ are drawn from a generic distribution, which, in the case of a fixed sample size $N$, can equivalently be expressed in terms of the underlying latent intensities as $p(\theta)$.
However, any analytical description of component systems ultimately requires some additional knowledge of these latent variates.
We discussed already in previous sections that while the latent distribution cannot be measured explicitly, it can still be inferred indirectly from empirical measurements.

\begin{figure}[!ht]
  \centering
  \includegraphics[width=\textwidth,height=.58\textheight,keepaspectratio]{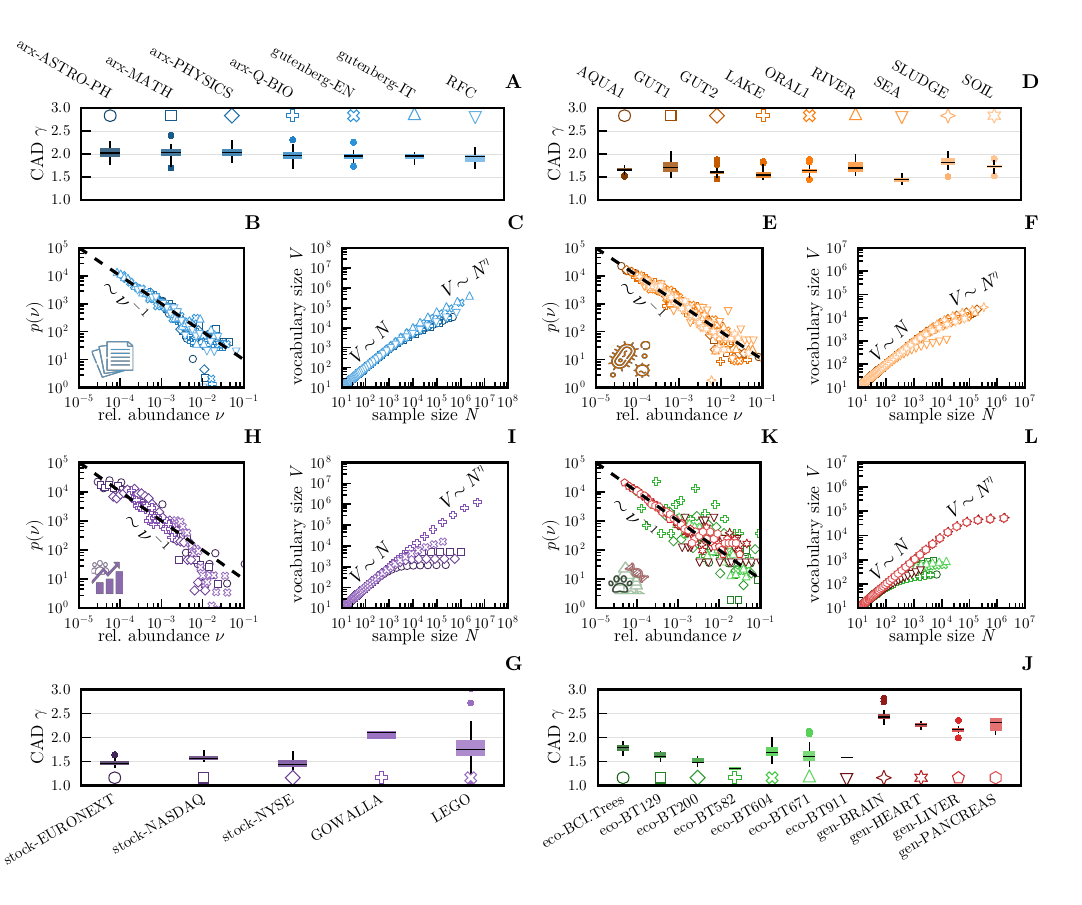}
  \caption{\textbf{Emergent scaling laws in complex component systems.}
  Canonical scaling laws, specifically Zipf's and Heaps' law, are observed across a wide variety of complex systems, grouped here into five broad categories: linguistic (blue), microbial (orange), social activities (purple), ecological (green), and biological (red).
  (\textbf{A}, \textbf{D}, \textbf{G}, \textbf{J})~Exponents of component abundance distributions in all datasets that we analyzed.
  Box plots depict mean and variance of the exponent $\gamma$ across all samples obtained by fitting a heavy-tailed distribution (see~Methods).
  (\textbf{B}, \textbf{E}, \textbf{H}, \textbf{K})~Aggregated CADs for each category.
  The aggregation of all samples within a specific category is again heavy-tailed with the same exponent.
  Data have been rescaled accordingly to emphasize the universality of the underlying distribution (Methods).
  (\textbf{C}, \textbf{F}, \textbf{I}, \textbf{L})~Growth of vocabulary $V(N)$ versus aggregated sample sizes $N$, obtained by repeated concatenation of samples within a category.
  Growth shows Heaps' law, with an intermediate sublinear regime consistent with the exponent of the CAD.
  Note that most datasets never reach saturation meaning that the total sampling effort $N$ is (much) smaller than the characteristic scale of the system $\varphi$.
  Exceptions to this are visible in financial and genetic data as in these systems (almost) all components are known and extracted.}
  \label{fig:cad_heaps}
\end{figure}

One of the most robust observations across complex systems is that component counts often display heavy-tailed distributions (\cite{newman2005power,kendal2011tweedie,mazzolini2018statistics}, Fig.~\ref{fig:cad_heaps}).
Although observed counts and latent frequencies are not identical objects, under the Poisson-mixture model, heavy-tailed latent frequencies naturally generate heavy-tailed observed counts.

Thus, heavy-tailed distributions of empirical relative counts can be interpreted as signatures of heavy-tailed latent intensity distributions.
Motivated by this and extensive analyses of dozens of datasets across fields (see~Methods and Supplementary Information 1), we consider latent intensities $\theta_{ik}$ to be drawn independently from a generic heavy-tailed distribution,
\begin{equation}
    \label{eq:powerlaw}
    p(\theta) \sim G(\theta) \theta^{-\gamma}, \qquad \theta \geq \theta_{\min},
\end{equation}
with exponent $\gamma > 1$ and a $G(\theta)$ a slowly varying function~\cite{bingham1987regular}.
Notice that Taylor's law implies that variation in relative counts must be finite, which indicates that the latent distribution must have finite moments.
Thus, whereas the latent distribution may be heavy-tailed across decades, these distributions must eventually temper or truncate.
This constraint is effectively imposed by the fact that all realistic systems are finite.
Finally, we assume that the latent distribution is fixed across observations (Supplementary Information 3), while the latent variables themselves may vary from one observation to another.

We tested this assumption by considering the \emph{component abundance distribution} (CAD), defined as the distribution of component abundances within a single observation, $p_k(\nu)$.
Across all datasets, the CAD displays asymptotic heavy-tailed behavior.
Moreover, within each dataset, the tail exponents estimated from different observations, $\gamma_k$, fluctuate around a dataset-specific typical value, $\gamma^*$ (Fig.~\ref{fig:cad_heaps}).
Therefore, the observed abundance distributions are consistent with the existence of a heavy-tailed distribution of component latent rates that is shared across observations.

Trivially, Zipf's law is obtained directly from the statistical properties of order statistics (Supplementary Information 6), so that
\begin{equation}
  \label{eq:zipf}
  \nu_{ik} \sim r_{ik}^{-\zeta},
\end{equation}
where $r_{ik}$ is the rank of the $i$th component in sample $k$, and $\zeta = 1 / (\gamma - 1)$ the corresponding Zipf exponent.

Notice that this within-dataset concentration of $\gamma_k$ is central to our interpretation. 
Zipf's law (and its connection with Heaps', see below) follows once a heavy-tailed abundance distribution is assumed.
What must be established empirically is that different samples from the same dataset are compatible with a common latent abundance structure.
The observed fluctuations of $\gamma_k$ around a dataset-specific $\gamma^*$ support this assumption.

\startfullwidthfigure
\finishfullwidthfigure

\subsection*{Limits of observation and scaling in vocabulary growth}\label{sec:heaps}
As complex component systems typically contain a staggering amount of components, observations are almost never sufficient to fully resolve a systems.
More formally, this means that the sample size $N$ is much smaller than the \emph{characteristic scale} $\varphi$ inherent to the system.
In this context, $\varphi$ may be interpreted as the minimum sample size that is required in order to expect every component to be observed at least once.

The existence of such a scale naturally separates the process of observation into three distinct regimes.
These regimes can be identified by considering the \emph{vocabulary size} $V(N)$, i.e.~the number of distinct components observed in a sample of size $N$ (Fig.~\ref{fig:illustration}C).
When $N \ll \varphi$, almost every observed component has not yet been observed, and hence vocabularies grow linearly~\cite{eliazar2011growth,mazzolini2018heaps}, i.e.
\begin{equation}
  V(N) \propto N  
\end{equation}
In contrast, when $N \gg \varphi$, all components have been detected and the vocabulary size must saturate, yielding
\begin{equation}
  V(N) \rightarrow S  
\end{equation}
where $S$ is the total number of distinct components in the system.
Note that $S$ is generally both unknown and extremely large, so the saturated regime wherein $N \gg \varphi$ is unlikely to be observed in most empirical systems (exceptions occur in systems such as financial portfolios, or genetic sequences, where the effective component space is known~\cite{nurk2022complete}, see~Fig.~\ref{fig:cad_heaps}).
Between these two extremes lies an intermediate regime wherein a non-trivial scaling law emerges.

More specifically, in the case of a heavy-tailed CAD, the progressive resolution of increasingly rare components gives rise to a scaling relation between the vocabulary size $V(N)$ and the sample size $N$.
Particularly, we find Heaps' law~\cite{herdan1958relation,heaps1978information}, which reads (Supplementary Information 7)
\begin{equation}
  V(N) \sim N^\eta,
\end{equation}
where $\eta = \gamma - 1$, in accordance to literature~\cite{eliazar2011growth}.

This emphasizes that the apparent scale-free growth of vocabulary sizes is not only a fundamental property of the system, but additionally reflects the progressive unveiling of a heterogeneous population under sampling with finite resolution.
The scaling exponent is therefore determined jointly by the structure of the latent distribution and by the constraints imposed by the observation process.
When observations are of sufficient lengths to resolve a significant fraction of components that comprise the system, Heaps' law thus emerges naturally as a consequence of sampling from a heavy-tailed distribution of relative abundances.
Moreover, Heaps' law, as was Taylor's law, is also transient.
Eventually, vocabulary growth will inevitably stagnate and any trace of scaling disappears completely.

\section*{Discussion}
\label{sec:discussion}
In the past decades most efforts to explain scaling laws in complex systems focused on generative mechanisms that reproduce individual laws within specific domains.
Here, we instead show that multiple canonical scaling laws emerge simultaneously from generic statistical properties of both the system and the observation process.
From this perspective, these laws can arise as statistical baselines inherent to sampling of heterogeneous systems, without invoking any generative mechanisms.
Moreover, the scaling relations observed in empirical systems should often be understood as transient regimes reflecting finite sampling instead of fundamental asymptotic laws.

Although statistical convergence may possibly lead one to think that the laws themselves are uninteresting, we instead argue the opposite.
Similar to the central limit theorem, it is not Gaussian distributions that are interesting \emph{per se}, but deviations from them or the actual values of the moments of the distributions, that are.

In this view, the GTEx data provide an informative boundary case.
Unlike most datasets, gene-expression data do not reproduce the component-wise quadratic coefficient with the same accuracy across independent splits, and a generic power-law fit provides a better empirical description.
This suggests that the effective latent variability in gene expression is shaped by additional biological and technical structure, such as donor-to-donor variability, batch effects, library-size effects, tissue composition, or cell-type heterogeneity.
Rather than invalidating the sampling baseline, this deviation illustrates how departures from the null model can identify systems where domain-specific mechanisms or covariates remain visible at the level of component counts.

Similarly, the analysis on CADs reveal that distribution exponents are very stable between samples belonging to the same dataset, supporting the idea of a common latent distribution.

For example, frequencies in texts tend to exhibit a CAD with exponent $\gamma \approx 2$ (thus, Zipf with exponent $\zeta \approx 1$~\cite{zipf1942unity,newman2005power,petersen2012languages}), and analyses on microbial OTUs (including the one presented here) reveals exponents $\gamma \approx 1.5\textrm{--}2$ across a wide variety of environments ranging from the human gut to glacier biomes.
These results strongly suggest constraints beyond mere heavy-tailed latent structures, similar to those put forward when discussing Zipf's law~\cite{cancho2003least}.

Our results thus posit that if scaling laws arise generically from heterogeneous sampling, then the central scientific questions should shift from explaining the presence of these laws to explaining why specific systems realize particular exponent values instead.
In this view, our results do not eliminate the need for mechanisms.
Instead, they identify which part of the observed scaling can already be expected from heterogeneous sampling, thereby clarifying what specific mechanisms can explain, such as system-specific values of the latent distributions and exponents.

Despite its simplicity, our framework relies heavily on the assumption that counts arise from observations of component systems whose rates admit an effective representation as variates with heavy-tailed distributions.
Whereas all empirical data that we examined did exhibit heavy-tailed CADs, we did not intend to explain the omnipresence of these distributions in itself.
Therefore, why exactly such distributions occur in many complex systems remains an open question whose answers lie beyond the scope of this work.
Nevertheless, many routes to the emergence of power laws in complex systems have been proposed, relying on a variety of mechanisms, from stochastic evolution to optimization processes~\cite{carlson1999highly,carlson2000highly,simon1955class,frieden2005power,mitzenmacher2004brief}.
For example, self-organized criticality may underlie scale invariance, which in turn signifies the presence of power laws~\cite{bak1987selforganized,sornette1989self,vespignani1998how,sornette1998discrete}.
Additionally, the same statistical convergence that may predicate the shapes of abundance fluctuations (see, e.g., \cite{kendal1995probabilistic,kendal2011taylors,grilli2020macroecological}, and Supplementary Information 4.4) may again bring about scale invariance in latent distributions.

Altogether, our findings suggest that progress in understanding complex component systems will come not from cataloguing mechanisms that generate scaling laws but from figuring out how systems select their parameters or depart from statistical convergence.

\section*{Methods}
\label{mm:methods}

\subsubsection*{Fitting heavy-tailed distributions}
To establish component counts following heavy-tailed data we performed extensive statistical analyses (Supplementary Information 1).
Specifically, for each sample across all datasets, we follow the methodology of~Ref.~\cite{clauset2009powerlaw} to test consistency with a distribution from the family of Pareto distributions.
Note that our aim is not to determine the globally best-fitting model, but to verify whether the tail is compatible with a heavy-tailed distribution, of which most are Pareto-like distributions (for an overview of all candidate distributions, see~Supplementary Table S5).

For each sample, we empirically verify which candidate of the Pareto family of distributions best describes the data, and we subsequently perform a goodness-of-fit test to compute a $p$-value.
This value quantifies the probability (under the fitted model) of observing a discrepancy at least as large as that measured in the data.
Following~Ref.~\cite{clauset2009powerlaw}, we reject the heavy-tailed hypothesis when $p \leq 0.1$.
If the hypothesis is not rejected, the procedure yields estimates of both the exponent $\gamma$ and the threshold $\varepsilon$, above which the CAD is considered to follow the distribution.
We then restrict the data to $\nu \geq \varepsilon$ and fit alternative candidate distributions.
Model selection is carried out using the Akaike Information Criterion (AIC)~\cite{burnham2004multimodel},
\begin{equation}
  \mathrm{AIC} = 2u + 2\,\mathcal{L}(\boldsymbol{\nu}),
\end{equation}
where $\mathcal{L}(\boldsymbol{\nu}) = - \sum_{i=1}^{S^\prime} \log F(\nu_i; \boldsymbol{\Omega})$ is the negative log-likelihood and $u$ is the number of parameters $\boldsymbol{\Omega}$ (e.g., for a standard ParetoI distribution, $u=2$).
The model with the smallest AIC, $AIC_{\min} = \min_w AIC_w$, is selected as the most plausible description of the tail.
If one desires, one may compute relative likelihoods to directly compare models
\begin{equation}
  \ell_w = \exp \left[ -\tfrac{1}{2} \left( AIC_w - AIC_{\min} \right) \right],
\end{equation}
and the associated normalized weights
\begin{equation}
  p_w = \operatorname{Prob} \left[ w \mid \boldsymbol{\nu} \right] = \ell_w \big/ \sum_w \ell_w,
\end{equation}
which quantify the support for each candidate model~\cite{gerlach2013stochastic}.
That is, for each of the models $w$ it gives a probability of that model describing the data among a set of candidate models.
Typically, for heavy-tailed data, distinct members of the Pareto family fit almost equally well, thus comparisons should only be made between the best heavy-tailed distribution and those with tails that decay exponentially.
We found that in most datasets we analysed component abundances are best described by a heavy-tailed distribution in over 80\% of them~(see~Supplementary Table S4), supporting our assumption that there exists latent variates that follow~$\theta \sim \theta^{-\gamma}$ asymptotically (Supplementary Information 3).

\subsubsection*{Comparing histograms of heavy-tailed component counts}
In Fig.~\ref{fig:cad_heaps} we report the rescaled CAD for each dataset.
Although the statistical analysis is performed at the level of individual samples (after filtering as in~Supplementary Information 1 and retaining only those with heavy-tailed behavior), the figure shows the aggregated CAD obtained by pooling counts across samples.
Aggregation generally produces a mixture distribution.
However, since individual samples exhibit power-law tails with exponents fluctuating around a characteristic value $\gamma^*$ (Supplementary Table S4), the aggregated distribution retains a power-law tail with exponent approximately $\gamma^*$.
To facilitate comparison across datasets for visualization purposes, histograms of relative frequencies $\nu$ are rescaled as $\nu \rightarrow \nu^{1/\gamma^*}$, which highlights the common scaling behavior and leads to CADs with exponent $\hat{\gamma} \approx 1$, as indicated in~Fig.~\ref{fig:cad_heaps}.

\section*{Data availability}
All datasets analyzed in this study are publicly available and are described in detail in Supplementary Information 1.

\section*{Code availability}
Code required to reproduce the results is available at \url{https://github.com/luca430/Meris}.

\section*{Acknowledgements}
We thank H. J. Jensen for insightful discussions and suggestions.

\section*{Funding}
L.A., J.N., and M.D.D. disclose support for the research of this work from the Human Frontier Science Program Organization (HFSP grant RGY0064/2022).

\section*{Author contributions}
L.A. and J.N. conceived the study, performed the analysis, and wrote the manuscript. M.D.D. supervised the research, provided critical feedback, and contributed to revising the manuscript. All authors discussed the results and approved the final version of the manuscript.

\section*{Competing interests}
The authors declare no competing interests.

\clearpage
\onecolumn
\newgeometry{margin=1in}
\fontsize{12pt}{14.5pt}\selectfont
\setcounter{figure}{0}
\setcounter{table}{0}
\setcounter{equation}{0}
\setcounter{section}{0}
\renewcommand{\thefigure}{S\arabic{figure}}
\renewcommand{\thetable}{S\arabic{table}}
\renewcommand{\theequation}{S\arabic{equation}}
\renewcommand{\thesection}{S\arabic{section}}
\renewcommand{\thesubsection}{S\arabic{section}.\arabic{subsection}}
\captionsetup{font=small,labelfont=bf,labelsep=period}

\raggedbottom
\vbadness=10000

\begin{titlepage}
\begin{center}
  {\Large\bfseries Supporting Information for}\\[1.5\baselineskip]
  {\Large\bfseries Scaling laws in complex component systems as consequences of heterogeneous sampling}\\[1.5\baselineskip]
  Luca Allegri, Johannes Nauta, and Manlio De Domenico\\[0.5\baselineskip]
  Correspondence: luca.allegri.1@phd.unipd.it, johannes.nauta@unipd.it, or manlio.dedomenico@unipd.it
\end{center}

\vspace{1.0\baselineskip}

\noindent\textbf{Preprint note.}
This supplementary material accompanies the original version submitted to Nature Communications and has not yet undergone peer review.

\vspace{1.0\baselineskip}

\bigskip
\noindent This PDF file includes:

\begin{itemize}
  \item Supplementary Information text
  \item Figs. S1 to S6
  \item Tables S1 to S5
  \item Supplementary references
\end{itemize}
\end{titlepage}

\setcounter{page}{2}

\section*{Supplementary Information}\label{sm}

\section{Detailed description of data}\label{sm:data}
In this work, we analyze datasets that can be represented in the form of component counts per sample.
The data span fundamentally different scientific domains, which we grouped into five macrocategories: linguistics, microbial systems, human activities, ecology, and biology.
These systems were chosen as data is readily available, however other systems will likely exhibit the same patterns.
A summary description with references for all datasets is provided in~Table~\ref{sm:tab:data}.
We made both the raw data and the scripts to download them available.

Within each dataset, we use the term \emph{class} to refer to a specific subdivision or partition of the data, such as different projects, environments, or tissues, depending on the context.

Classes with few samples or samples with few components were removed as fitting heavy-tailed distributions on a limited number of data entries that spans only a few decades is not practical.
As such, these criteria, reported in~Table~\ref{sm:tab:filter}, are typically large and chosen empirically, as they need to take into account the typical scales of the system.
We applied an additional filtering step by retaining only those classes containing at least 30 samples, each with as many distinct components as possible (Table~\ref{sm:tab:filter}, note that an exception is made for the BCI tree dataset where we set the minimum number of distinct components to 50, as most samples contain fewer than 100 different species).
No additional preprocessing was performed, to leave the data as intact as possible and avoid potential undesired biases.
We verified that changes in the preprocessing scheme do not alter our results regarding the canonical laws.

\paragraph{Texts}
All datasets in the linguistic category were constructed by tokenizing raw text, excluding numbers and symbols.
Lemmatization was not performed because it is not expected to significantly affect the overall component distribution given the large number of distinct components.
Texts were obtained from three primary sources: the Gutenberg Project~\cite{project_gutenberg}, arXiv~\cite{arxiv}, and the RFC document collection~\cite{rfc_archive}.
Specifically, we arbitrarily considered the first 500 Italian and English books from the Gutenberg Project ordered by their internal ID.
For \texttt{arXiv} entries, we used the \texttt{arxiv} API in \texttt{Python} to download the first 100 papers published in 16 different categories [\url{https://arxiv.org/}] from 2025-06-01 to 2025-12-31.
The categories selected are: \textit{q-bio.NC}, \textit{q-bio.BM}, \textit{q-bio.PE}, \textit{q-bio.GN}, \textit{astro-ph.GA}, \textit{astro-ph.CO}, \textit{astro-ph.EP}, \textit{astro-ph.SR}, \textit{physics.soc-ph}, \textit{physics.acc-ph}, \textit{physics.ao-ph}, \textit{physics.plasm-ph}, \textit{math.AG}, \textit{math.AT}, \textit{math.AP}, \textit{math.CT}.
These are later grouped into four groups: \textit{q-bio}, \textit{astro-ph}, \textit{physics}, and \textit{math}.

\paragraph{Microbial systems}
Microbial data originate from the EBI Metagenomics database~\cite{mitchell2018ebi}.
However, we used the revisioned version provided by~\cite{grilli2020macroecological}, in which data from multiple independent projects were collected and data was parsed and standardized.
These data are readily available at \url{https://github.com/jacopogrilli/lawsdiv} and comprise a vast set of environments: \emph{oral cavity}, \emph{gut}, \emph{vaginal}, \emph{soil}, \emph{feces}, \emph{human skin}, \emph{glacier}, \emph{sea water}, \emph{river water}, \emph{lake water}, \emph{aquatic thermal water}, and \emph{sludge}.
Some of these environments appear multiple times as they reflect data collected from different projects. 
Some environments were automatically excluded by our filtering procedure.

\paragraph{Human activities}
Human activity data include human mobility records from the Gowalla project~\cite{cho2011friendship}, LEGO sets~\cite{rebrickable}, and daily financial stock volume exchange data from three large financial markets available at~\cite{yahoo_finance}.
For the LEGO dataset, we restricted our analysis to the 100 themes that contain the largest number of distinct sets.
Daily Yahoo Finance volume data for 2025 were downloaded using the \texttt{yfinance} package in Python.
We considered three major financial markets: NYSE, NASDAQ, and Euronext.
For the Euronext market, the three largest were selected, which are Paris (Paris Bourse), Amsterdam (AEX), and Milan (Borsa Italiana).

\paragraph{Ecological systems}
Ecological data was obtained from two main sources: the Barro Colorado Island (BCI) trees dataset~\cite{condit2012barro} and the BioTIME project~\cite{dornelas2018biotime}.
In the BCI trees dataset, individual trees are recorded within labelled quadrats (patches).
We found that counts per quadrat are typically low, making fitting heavy-tailed distributions practically impossible.
Therefore, we randomly aggregated quadrats in pairs to form larger spatial units, thereby increasing sampling depth which reduced the number of zeros within a single patch.
The BioTIME database contains data from several hundred independent projects.
After applying our filtering criteria, only the classes reported in this work were retained as many projects either consider very few species or have very few counts from a restricted spatial or temporal grid.

\paragraph{Genetic data}
Genetic data were obtained from the GTEx project~\cite{gtex2020atlas}.
For this study, we selected four tissues: heart, brain, liver, and pancreas.

\section{Sampling complex component systems}\label{sm:sampling}
In this work we consider samples composed of discrete components, each appearing a certain number of times. 
Formally, a sample $k$ is represented by the collection of counts $\{n_{1k}, \, n_{2k}, \, \dots, \, n_{Vk} \}$, where $V$ denotes the number of distinct components in the system.
We imagine a component system as an urn containing a number $M$ of elements, which is so large that can be considered effectively infinite.
A sample $k$ is then composed of $N_k = \sum_{i=1}^V n_{ik} \ll M$ total components.
We suppose that the generation of a sample is a discrete process in which an observer blindly picks components one by one (Fig.~1 in the main text).
In this view, focusing on component $i$, we can assign a random variable $X_i^{(t)}$ which is $1$ if observation $l$ results in species $i$, and it is $0$ otherwise.
Then, after $N_k$ observation there will be a sequence $\{X_{ik}^{(1)}, X_{ik}^{(2)}, \dots, X_{ik}^{(N_k)}\}$ assigned to species $i$ such that
\begin{equation}
    n_{ik} = \sum_{t=1}^{N_k}X_{ik}^{(t)}.
  \end{equation}

  In statistics, a sequence of random variables $\{Y_1, Y_2,\dots,Y_q\}$ is said to be exchangeable if the joint probability
\begin{equation}
    \label{sm:eq:exchangeable}
    \mathcal{P}(\{Y_1, Y_2,\dots,Y_q\}) = \mathcal{P}(\{Y_{\pi(1)}, Y_{\pi(2)},\dots,Y_{\pi(q)}\})
\end{equation}
where $\pi$ represents a permutation of the indices.
Depending on the observer strategy, the joint probability of a sequence $\{X_{ik}^{(1)}, X_{ik}^{(2)}, \dots, X_{ik}^{(N_k)}\}$ can or cannot be exchangeable.
For example, consider the case of a book in which words are the components.
Focusing on the word ``cat'', when components are gathered sequentially, the sequence $\{\text{``the''}, \text{``cat''} \}$ is much more probable than the sequence $\{\text{``cat''}, \text{``the''} \}$, meaning that sequential sampling of words creates non exchangeable sequences.
If instead the sampling is done more in the spirit of drawing balls from a urn, then the two sequences will have the same probability, thus they would be exchangeable.

De Finetti's theorem states that infinitely long exchangeable sequences of Bernoulli random variables are conditionally independent and identically distributed given the exchangeable sigma-algebra (i.e., the sigma-algebra consisting of events that are measurable with respect to a sequence and invariant under finite permutations of its indices).
In other words, given a latent variable $\lambda_k$,
\begin{equation}
    \mathcal{P}(X_{ik}^{(1)},\dots, X_{ik}^{(N_k)} \vert \lambda_k) = \prod_{t=1}^{N_k}\mathcal{P}_t(X_{ik}^{(t)}).
\end{equation}

Notice that the theorem considers \emph{infinite} sequences of Bernoulli variables, but in reality each sequence an observer may collect will always be finite.
In~\cite{diaconis1980finite} the authors extend de Finetti's result to long but finite sequences showing that
\begin{equation}
    \Vert\mathcal{P}_{\text{exch}}^N - \mathcal{P}_{\text{iid mix}}^N \Vert \le \frac{N(N-1)}{2M}
\end{equation}
where $\mathcal{P}_{\text{exch}}^N$ is the joint probability of an exchangeable sequence of length $N$, $\mathcal{P}_{\text{iid}}^N$ is the probability of $N$ conditionally iid Bernoulli variables and $M$ is the maximum length of a sequence.

Thus, for unordered component-count observations that can be approximated as exchangeable, De Finetti-type results motivate a latent-variable representation.

\section{Heavy-tailed distributions of latent variables}\label{sm:latentdistribution}
De Finetti's theorem, and its finite-sample extensions, provide a more rigorous justification for the introduction of latent variables in component systems.
Specifically, the theorem states that any exchangeable sequence of Bernoulli random variables can be represented as conditionally independent given a latent parameter.
In this representation, the latent parameter determines the probability of observing the component, and the observed counts arise as sums of independent Bernoulli trials conditioned on that parameter.
Importantly, the latent parameter is itself a random variable drawn from a common distribution, which we refer to as the \emph{latent distribution}.
This distribution captures the variability of component frequencies across different samples.

To illustrate this idea, consider an ensemble of urns, each containing red and blue balls but with different fractions of red balls.
Let $p_k$ denote the fraction of red balls in urn $k$.
If one repeatedly draws balls from a given urn, the resulting sequence of observations is conditionally independent given $p_k$.
However, across different urns the value of $p_k$ varies, and can therefore be viewed as a random draw from an underlying distribution $p(p_k)$.
In this interpretation, the observed counts reflect two sources of randomness: the sampling process within each urn, and the variability of the latent parameter across urns.

This framework naturally generalizes to systems with many components.
Each sample is characterized by a vector $\boldsymbol{p}_k$ of latent intensities, where each entry is drawn from a common latent distribution that describes the system as a whole, so that $p_{ik} \sim \rho(p)$.
Individual samples therefore represent different realizations of the same underlying statistical structure, analogous to the concept of ensembles in statistical physics.

This intuition is empirically verified by comparing the CADs of different samples belonging to the same dataset.
In Fig.~3 in the main text the boxplots show that the exponents computed for the tails of each CAD are very similar, suggesting that each sample is indeed composed of components distributed according to a random vector of latent intensities.

\section{Taylor's law as a consequence of sample variability}\label{sm:taylor}
In the main text we argued that Taylor's law emerges universally as a result from the law of total variance.
Here, we provide a more detailed derivation.

Taylor's law, as listed in~Eq.~(2) in the main text, states that the variance of counts of component $i$ in observation $k$ scales with their mean as a power law with exponent $b$, i.e.
\begin{equation}
  \operatorname{Var}\!\left[n_{i;k}\right] \propto \operatorname{E}\!\left[n_{i;k}\right]^b
\end{equation}
As such, Taylor's law describes components whose fluctuations grow as their expected number of counts increases.
Typically, the exponent $b$ is found to be in the range between $1$ and $2$ across a diverse set of systems~\cite{eisler2008fluctuation}.
Similar to previous work on this topic, this range should be considered dynamic or transient~\cite{james2018zipfs}, as not all components display a similar approximate unified exponent.
Instead, we highlight here that $b$ actually depends on the rate of occurrence of a component.
Common components will eventually converge to exhibit exponents $b \rightarrow 2$ as sample sizes increase.
In contrast, statistics of rare components, due to their inherent significant probabilities of not being sampled, will be dominated by sampling noise instead of genuine system heterogeneity.
This will inevitably lead to Taylor's exponents $b \rightarrow 1$.

To show this, we focus here on a set composed of $K$ observations or samples, each with a potentially distinct size $N_k$, $k=1\ldots, K$.
We focus on the case where an observer is interested in estimating the expected number of counts in a sample of size $N_k$, which they infer through their observations and computing sample statistics,
\begin{equation}
  \mu_i = \frac{1}{K} \sum_k n_{i;k},\qquad \sigma_i^2 = \frac{1}{K} \sum_k (n_{i;k} - \mu_i)^2
\end{equation}
Subsequently, an observer may plot the sample variances against the means and when $\sigma_i^2 \propto \mu_i^b$ they may invoke Taylor's law.

From a theoretical perspective, the sample moments represent the true moments of the sampling process assuming these fall within the domain of the central limit theorem.
That is, when $K \rightarrow \infty$ one would expect that the sample moments converge to their true values, and thus they are representative of the genuine underlying mean and variance of component counts.
In practice, the number of samples in each observation, $N_k$, is typically large, yet the total number of distinct samples, $K$, does not have to be.
This will inevitably introduce deviations from Taylor's law (assuming that it holds), and thus one should expect Taylor's law to hold not strictly, but instead to hold approximately.
Theoretically, however, Taylor's law holds strictly~(see, e.g., Fig.~1a in the main text), as we shall now show.

Following the minimal model introduced in the main text, which assumes that counts are generated by a Poisson process with random rates $\lambda_{i;k}$, yields the law of total variance
\begin{equation}\label{sm:eq:total_variance}
  \operatorname{Var}_k\!\left[n_i\right] =
  \operatorname{Var}_k\!\big[\operatorname{E}_k\!\left[n_i | \lambda_i\right] \big]
  + \operatorname{E}_k\!\big[ \operatorname{Var}_k\!\left[n_i | \lambda_i\right] \big],
\end{equation}
where both the expectation and variance are taken \emph{across} samples, as indicated by $k$.
That is, there exists some rate with expectation $\operatorname{E}\!\left[\lambda_i\right] := \operatorname{E}_k\!\left[\lambda_{i;k}\right]$ and variance $\operatorname{Var}\!\left[\lambda_i\right] := \operatorname{Var}_k\!\left[\lambda_{i;k}\right]$, which is straightforwardly evaluated under the assumption that samples follow from a Poisson process where $n_{i;k} \sim \operatorname{Poisson}(\lambda_{i;k})$.
Considering a Poisson process, Eq.~\ref{sm:eq:total_variance} reads
\begin{equation}
\label{sm:eq:var}
  \operatorname{Var}\!\left[n_i\right] = \operatorname{Var}\!\left[\lambda_i\right] + \operatorname{E}\!\left[\lambda_i\right]
\end{equation}
Now, as any rate must scale linearly with system size, we let the rates $\lambda_{i;k}$ depend on some latent intensity $\theta_{i;k} \geq 0$.
One way to see this is to realize that $\operatorname{E}\!\left[n_i\right] = \operatorname{E}\!\left[\lambda_i\right] \propto N$: the total counts of some component should increase when observation lengths increase.
We let these intensities follow a \emph{latent distribution}, $\theta_{i;k} \sim p(\theta)$.
Importantly, intensities must have finite first and second moments, a condition related to the fact that all component systems are finite (see also~Supplementary Note~\ref{sm:latentdistribution}).
This implies that this \emph{distribution} does not vary substantially between subsequent observations.
If the distribution is of a particular form with a set of parameters, both the form and parameters are assumed to not vary significantly between subsequent observations.

Under the latent variate assumption, each component has an intrinsic rate $\lambda_{i;k} = N_k \theta_{i;k}$, which crucially depends on the observation length $N_k$.
This is intuitive, as larger samples should reveal more components by design.
Despite this dependence, we consider here the simpler case where $N$ is fixed across observations to highlight the emergence of Taylor's law.
We could relax this assumption (which we do below~Supplementary Note~\ref{sm:samplingheterogeneity}), but note that this condition can always be met by downsampling counts to a common sample size $N$.
Intuitively, any regularity (i.e., any ``law'') should also \emph{not} depend on \emph{how} observations come to be --- it is the \emph{properties} of both the sampling and the system from which the scaling law emerges.
After all, if it mattered precisely how counts are obtained, it is likely that these laws had not been uncovered in the first place.

Now, as $\lambda_{i;k} = N \theta_{i;k}$ we may write $\operatorname{Var}\!\left[\lambda_i\right] = N^2 \operatorname{Var}\!\left[\theta_{i;k}\right]$ and $N = \operatorname{E}\!\left[n_i\right] / \operatorname{E}\!\left[\theta_{i;k}\right]$.
Thus, Eq.~\ref{sm:eq:var} reads
\begin{equation}
  \label{eq:sm:variancemean}
  \operatorname{Var}\!\left[n_i\right] = \Omega[\theta_{i}] \operatorname{E}\!\left[n_i\right]^2 + \operatorname{E}\!\left[n_i\right]
\end{equation}
with $\Omega[\theta_{i}] = \operatorname{Var}\!\left[\theta_{i}\right] / \operatorname{E}\!\left[\theta_{i}\right]^2$ the squared coefficient of variation of the latent intensities.
Again, we consider $\operatorname{E}\!\left[\theta_i\right] := \operatorname{E}_k\!\left[\theta_{i;k}\right]$, and similar for the variance.

Note that Eq.~\ref{eq:sm:variancemean} captures both the linear regime, wherein the variance is proportional to the mean (i.e., $b=1$), and the quadratic regime ($b=2$).
This raises the natural question of which regime one can expect to dominate.
First, if any genuine variation between observations is suppressed, then $\Omega[\theta_{i}] \rightarrow 0$, which leads to $b=1$.
This is the regime wherein sampling noise dominates system heterogeneity.
Second, if both $\operatorname{E}\!\left[n_i\right]$ and $\Omega[\theta_{i}]$ are sufficiently large then the quadratic term dominates, and we find Taylor's law with $b = 2$.
The transition between the two regimes happens at $\mu_i = \Omega[\theta_{i}]^{-1}$.

Importantly, as the quadratic coefficient $\Omega[\theta_{i}]$ depends on a specific component $i$, Eq.~\ref{eq:sm:variancemean} is component-dependent, meaning that each pair $(\mu_i, \sigma^2_i)$ is a point that belongs to the specific curve $y=\Omega_i x^2 + x$.
As the curve is only defined by the value of the quadratic coefficient, all components having the same coefficient of variation would appear as points of the same curve.
Since in general $\Omega \in \mathbb{R}$, it would be impossible to find two components $i$ and $j$ with $\Omega_i = \Omega_j$, but we can approximate this condition by binning the real axis and consider all species having $\Omega_i \in (\Omega_n^-, \Omega_n^+)$ to be part of the same effective quadratic curve.
This is exactly what is shown in~Fig.~2 in the main text.

\subsection{Test of the quadratic form}\label{sm:quadratic_test}
The quadratic form in~Eq.~(2) in the main text is tested in two complementary ways.
First, we compare the coefficients obtained by fitting the quadratic form $\sigma^2 = \mu + \Omega_{\mathrm{fit}}\mu^2$ on a train and a test set.
Second, we compare the Akaike information criterion (AIC) of the quadratic model with that of a power-law model, $\sigma^2 = a\mu^b$.

\begin{figure}[htbp]
  \centering
  \includegraphics[width=0.7\linewidth]{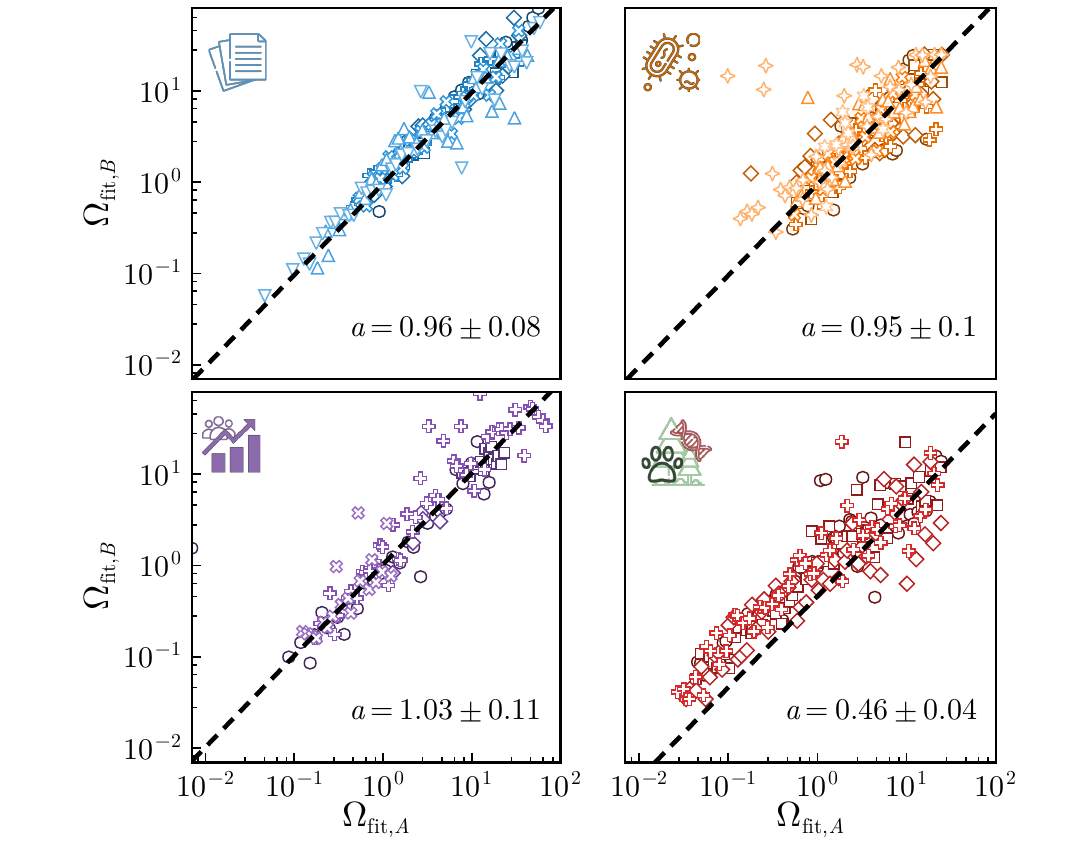}
  \caption{
    \textbf{Train--test comparison of the quadratic coefficient.}
    For each dataset, components are grouped according to similar empirical values of $\hat{\Omega}_i$ estimated on the train set.
    Within each group, the quadratic coefficient is then fitted independently on the train and test sets, giving the pairs $(\Omega^{\mathrm{fit}}_{\mathrm{train}},\Omega^{\mathrm{fit}}_{\mathrm{test}})$ shown in the scatter plot.
    The black dashed line is a fit of the form $y=ax$.
    Values of $a$ close to one indicate that the coefficient inferred from one subset predicts the coefficient measured on the other, supporting the component-wise validity of the quadratic form.
    Fits are performed on bin with at least $20$ components inside.
    }\label{sm:fig:tl-test1}
\end{figure}

For the first analysis, each dataset is divided into two independent subsets.
The first subset (train) is used to estimate the empirical coefficient $\hat{\Omega}_i=(\operatorname{Var}\!\left[n_i\right] - \operatorname{E}\!\left[n_i\right]) / \operatorname{E}\!\left[n_i\right]^2$ for each component $i$.
Components are then grouped according to similar values of $\hat{\Omega}_i$ and the quadratic model is fitted within each group.
The same component groups are then selected in the second subset (test), where the quadratic form is fitted again within each group.
This procedure avoids fitting and testing the same quantities on the same data while allowing to directly test the local nature of~Eq.~(2) in the main text.
Indeed, the quadratic coefficient is component-dependent: each component $i$ is expected to follow $\operatorname{Var}\!\left[n_i\right] = \operatorname{E}\!\left[n_i\right] + \Omega_i \operatorname{E}\!\left[n_i\right]^2$.
Therefore, if components are grouped by similar empirical coefficients, the fitted coefficient in the second subset should match that of the other, namely $\Omega^{\mathrm{fit}}_{train} \simeq \Omega^{\mathrm{fit}}_{test}$. The results are shown in Fig.~\ref{sm:fig:tl-test1}.

The second test consists of a direct AIC comparison between the quadratic model and the power-law model.
The full comparison is reported in~Table~\ref{sm:tab:taylor_model_comparison}.

From our analysis, in most datasets groups of components selected according to similar values of $\Omega$ in the training set retain comparable fitted coefficients in the test set, and the quadratic form is favored over a generic power-law fit.
An exception is given by genetic data for which the fitted coefficients are not reproduced with the same accuracy across independent splits, and the power-law model provides a better empirical description.
This deviation suggests that the effective coefficient $\Omega$ is not stable across samples, or that it depends systematically on the mean expression level.

A plausible interpretation is that gene-expression data contain additional structured sources of variability that are not fully captured by the minimal sampling model.
These may include donor-to-donor variability, batch effects, library-size effects, tissue composition, cell-type heterogeneity, or other technical and biological covariates.
Therefore, GTEx should be regarded not as a failure of the quadratic decomposition itself, but as a case in which the latent variability is shaped by additional structure.
In this sense, its deviation from the general trend is informative: it identifies gene expression as a system where the statistical baseline is modified by domain-specific sources of heterogeneity.

\subsection{The effect of rare and common components}\label{sm:common_rare}
Whereas the above derivation predicts Taylor's law to hold with $b=2$ for all components with $\Omega[\theta_{i}] \gg 1$, it is not yet clear under what conditions this is the case.
In fact, the above derivation assumed independent latent variates across observations, however this is often not the case as some components are inherently more common than others.
For example, in language, the word ``the'' accounts for approximately 7\% of all the words in a corpus regardless of the type of text.
Therefore, while the latent variate underlying its occurrences may vary (even though it is not the case, see Supplementary Note~\ref{sm:homogeneity}), it is typically among the largest in any observed collection of words.
This implies that the latent distributions may contain heterogeneity that is the product of genuine heterogeneity (which, for example, predicts the expected occurrence of the word ``the'') and continuous perturbations that manifest themselves as random fluctuations.
In other words, we may express the latent variates as
\begin{equation}
  \theta_{i;k} = \theta_i \xi_k  
\end{equation}
with $\xi_k \sim p_\xi(\xi)$ a random variate that is (approximately) equal for all components in the system --- i.e., it is a global system property.
Note that the tails of the distribution may safely be assumed to ``not be heavier'' than those of the latent distribution itself.

Within this context, Taylor's law may be readily interpreted as a transient scaling law.
That is, if we consider $\operatorname{E}\!\left[n_i\right] = N \theta_i \operatorname{E}\!\left[\xi_k\right]$ with $\operatorname{E}\!\left[\xi_k\right] < \infty$ finite, then components with $N \theta_i \gg 1$ will exhibit Taylor's law with $b=2$.
The transience comes from the fact that as $N$ keeps increasing, an observer will eventually have resolved all available structure as inevitably $N \theta_i \gg 1$ regardless of the value of $\theta_i > 0$.
In this regime, all components have been observed such that the law of large numbers dictates the convergence of the moments of their counts, and all components will exhibit $b=2$.

In practice, $N \theta_i \gg 1$ holds only for \emph{common} components.
In contrast, many \emph{rare} components will instead have $N \theta_i \ll 1$, meaning that the probability to observe them in a sample of size $N$ is in and of itself incredibly small.
In this case, sampling noise dominates, and the exponent will instead exhibit $b=1$.
As such, a natural transient regime from rare components, with $b=1$, to common components, with $b=2$, emerges when genuine heterogeneity overtakes sampling noise under the guise of the central limit theorem.

\begin{figure}[htbp]
  \centering
  \includegraphics[width=0.9\linewidth]{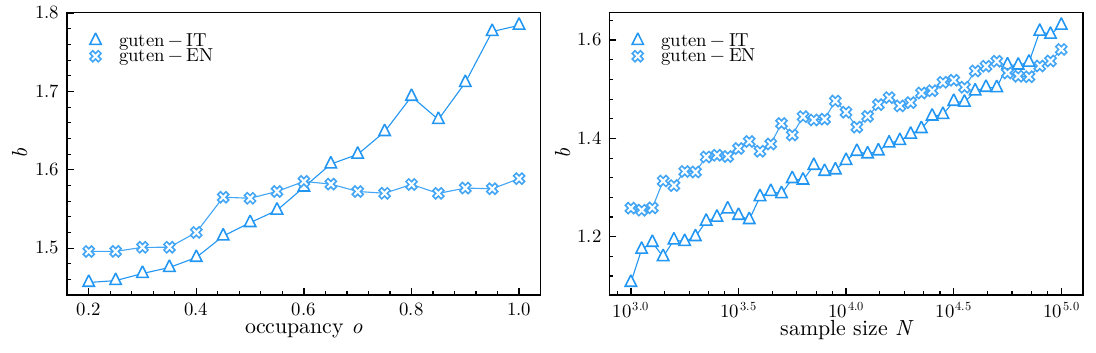}
  \caption{
  \textbf{Transition of Taylor's exponent.}
  (\textbf{Left}) The inclusion of rare components biases Taylor's exponent toward lower values.
  The plot is obtained by downsampling all samples to a fixed total number of reads, $N = 10^5$.
  (\textbf{Right}) Increasing the sample size mitigates the influence of rare components, shifting Taylor's exponent toward higher values.
  The plot is obtained by fixing the occupancy (rate of appearance across samples) to $o = 0.7$.
  }\label{sm:fig:transition}
\end{figure}

\subsection{Sampling heterogeneity}
\label{sm:samplingheterogeneity}
Above we assumed that one may safely assume that $N_k = N$.
Here, we show explicitly that relaxing this assumption does not change Taylor's law, which implies that, from a practical perspective, subsampling is not necessary for Taylor's law to emerge.

To see this, assume that each sample $k$ has some size $N_k \sim P(N_k)$.
Conditioned on the latent variates $\theta_{i;k}$ and $N_k$ we have $n_{i;k} \sim \operatorname{Poisson}(N_k \theta_{i;k})$.
Again applying the law of total variance, we obtain
\begin{equation}
  \operatorname{Var}\!\left[n_i\right] = \operatorname{E} \left[ \operatorname{Var}\!\left[n_i | \theta_{i;k}, N_k\right] \right]
  + \operatorname{Var} \left[ \operatorname{E}\!\left[n_i | \theta_{i;k}, N_k\right] \right]
\end{equation}
Similarly to the derivation above, we find that
\begin{subequations}
\begin{align}
  \operatorname{E} \left[ \operatorname{Var}\!\left[n_i | \theta_{i;k}, N_k\right] \right]
  &= \operatorname{E}\!\left[N_k \theta_{i;k}\right] = \operatorname{E}\!\left[N_k\right] \operatorname{E}\!\left[\theta_{i;k}\right] = \operatorname{E}\!\left[n_i\right] \\
  \intertext{and}
  \operatorname{Var} \left[ \operatorname{E}\!\left[n_i | \theta_{i;k}, N_k\right] \right]
  &= \operatorname{Var}\!\left[N_k \theta_{i;k}\right] \\
  &= \operatorname{E}\!\left[N_k\right]^2\operatorname{Var}\!\left[\theta_{i;k}\right] + \operatorname{Var}\!\left[N_k\right]\operatorname{E}\!\left[\theta_{i;k}\right]^2 + \operatorname{Var}\!\left[N_k\right]\operatorname{Var}\!\left[\theta_{i;k}\right]
\end{align}
\end{subequations}
Define the coefficients of variation,
\begin{equation}
  c_N = \frac{\operatorname{Var}\!\left[N_k\right]}{\operatorname{E}\!\left[N_k\right]^2}, \qquad c_\theta = \frac{\operatorname{Var}\!\left[\theta_{i;k}\right]}{\operatorname{E}\!\left[\theta_{i;k}\right]^2}  
\end{equation}
which, together with the fact that $\operatorname{E}\!\left[n_i\right] = \operatorname{E}\!\left[N_k\right]\operatorname{E}\!\left[\theta_{i;k}\right]$, yields
\begin{subequations}
  \label{eq:coeffsofvariation}
  \begin{align}
    \operatorname{E}\!\left[N_k\right]^2\operatorname{Var}\!\left[\theta_{i;k}\right] &= c_\theta \operatorname{E}\!\left[n_i\right]^2 \\
    \operatorname{Var}\!\left[N_k\right]\operatorname{E}\!\left[\theta_{i;k}\right]^2 &= c_N \operatorname{E}\!\left[n_i\right]^2 \\
    \operatorname{Var}\!\left[N_k\right]\operatorname{Var}\!\left[\theta_{i;k}\right] &= c_N c_\theta \operatorname{E}\!\left[n_i\right]^2
  \end{align}
\end{subequations}
which gives us the expression above to resolve fo
\begin{equation}
  \operatorname{Var} \left[ \operatorname{E}\!\left[n_i | \theta_{i;k}, N_k\right] \right] =
  \operatorname{E}\!\left[n_i\right]^2 \left( c_\theta + c_N + c_N c_\theta \right) := c \operatorname{E}\!\left[n_i\right]^2
\end{equation}
Subsequently, we may straightforwardly write
\begin{equation}
  \label{eq:sm:variancemeansample}
  \operatorname{Var}\!\left[n_i\right] = \operatorname{E}\!\left[n_i\right] + c \operatorname{E}\!\left[n_i\right]^2
\end{equation}
which is again Taylor's law as in~Eq.~\ref{eq:sm:variancemean}, yet with another constant prefactor for the quadratic term.
Clearly, from Eq.~\ref{eq:coeffsofvariation} one may see that this term again tends to dominate for common components for which $N_k \theta_{i;k} \gg 1$, while the Poisson sampling noise dominates for rare components with $N_k \theta_{i;k} \ll 1$, as before.
This illustrates that finite fluctuations of sample sizes do not alter the summary behavior of Taylor's law.
Moreover, the transient behavior that was discussed above additionally remains.

\subsection{Rate homegeneity}\label{sm:homogeneity}
\begin{figure}[htbp]
  \centering
  \includegraphics[width=0.5\linewidth]{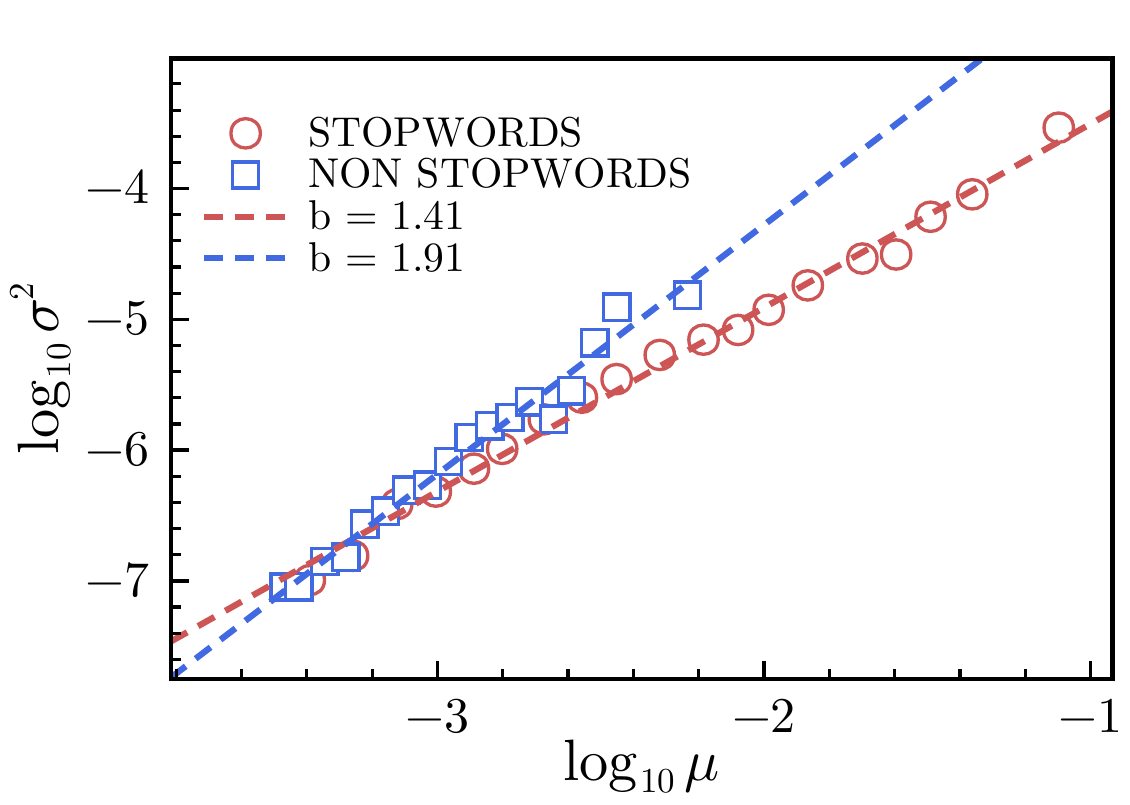}
  \caption{
  \textbf{Taylor's law of stopwords in texts.}
  When stop-words are considered alone, although being common components (Supplementary Note~\ref{sm:common_rare}) they show a Taylor's exponent $b \rightarrow 1$.
  This is because their rate of appearance in texts is approximately the same in all samples.
  Conversely, when stopwords are filtered out, common words display heterogeneity across samples, thus $b \approx 2$.
  Plot is made using the arXiv dataset and considering only components that appear in at least $95\%$ of samples. For visual reason, points have been binned in the x-axis and the average log-variance is taken within each bin.
  }\label{sm:fig:homogeneous}
\end{figure}

It is important to note that~Eq.~(2) in the main text predicts convergence to $b \to 2$ only when latent variables are heterogeneous across samples.
In contrast, when samples are highly homogeneous and component counts fluctuate only weakly around their mean, the sampling process approaches a purely Poisson (or Binomial) regime, yielding instead $b \to 1$.
For example, within the context of text data, when we restrict our analyses to stop-words only (e.g., ``the'', ``a'', etc.), Taylor's law exhibits an exponent closer to $1$.
The reason is that the occurrence rates of these words are approximately constant across documents due to their widespread use in natural language~(Fig.~\ref{sm:fig:homogeneous}).
Therefore, a substantial presence of components with homogeneous rates in a system can distort the log-log representation of Taylor's law, showing an exponent converging to values between 1 and 2.

This effect may also appear depending on the operative definition of a sample.
Recall that the original interpretation of Taylor's exponent refers to the level of species aggregation in ecological systems~\cite{taylor1961aggregation}, i.e.~higher values of $b$ imply that components have the tendency to aggregate in groups, while the opposite is true for lower values of $b$.
Extrapolating to general component systems, different levels of aggregation can create cases in which relative counts (and thus latent rates) of some components become very similar in all samples.
As a consequence, the level of component aggregation --- reflected in the exponent $b$ --- at one scale may appear different at a different level of coarse-graining.
This subsequently suggests a spontaneous breaking of the scale invariance property typical of component systems.
Whether this is truly the case, or how analyses may overcome this limitation to observations, is considered to be out of the scope of this work, however.

\subsection{Tweedie convergence and abundance fluctuations}\label{sm:tweedie}
Previously (see~Supplementary Note~\ref{sm:taylor}), we have shown that the simple sampling scheme proposed generates Taylor's law with exponent $b \in [1,2]$.
In this section, we show how the same scheme is enough to give a qualitative idea of the \emph{abundance fluctuation distribution} (AFD), which is the distribution of (relative) counts of a single component across samples (see, e.g., \cite{grilli2020macroecological,camacho-mateu2024sparse,camacho-mateu2024nonequilibrium}).
In particular, we show that an AFD that belongs to the so called family of~\emph{Tweedie distributions}~\cite{tweedie1947functions} emerges naturally from our framework.

Tweedie distributions are a special case of exponential dispersion models (EDMs)~\cite{jorgensen1987exponential,jorgensen1992exponential}, which are defined by probability densities of the form
\begin{equation}\label{sm:eq:edm}
  f(y \mid \vartheta, \phi)
  = a(y,\phi)\,\exp\!\left\{\frac{y\vartheta - \kappa(\vartheta)}{\phi}\right\},
\end{equation}
where $\vartheta$ is the canonical parameter, $\phi > 0$ is the dispersion parameter, and $\kappa(\vartheta)$ is the cumulant generating function (CGF).
The mean and variance are given by
\begin{equation}
\mu = \kappa'(\vartheta), \qquad \mathrm{Var}(Y) = \phi\,\kappa''(\vartheta).
\end{equation}
Tweedie distributions in particular are characterized by a power-law variance function of the form
\begin{equation}\label{sm:eq:tweedievar}
\mathrm{Var}(Y) = \phi\,\mu^{q},
\end{equation}
where $q \in \mathbb{R}$ is called the \emph{Tweedie index parameter}.
Different values of $q$ recover well-known distributions such as Gaussian ($q=0$), Poisson ($q=1$), Gamma ($q=2$), and inverse Gaussian ($q=3$).
Moreover, for $1<q<2$, the Tweedie distribution corresponds to a compound Poisson–Gamma distribution, representing sums of a Poisson-distributed number of Gamma-distributed increments.

Note that Eq.~\ref{sm:eq:tweedievar} is the same relation as Taylor's law, making Tweedie distributions of particular interest for our purposes.
Crucially, there exists a generalized central limit theorem for exponential dispersion models~\cite{morris1982natural,jorgensen1987exponential,kendal2011tweedie,kendal2011taylors}.
Just like the Gaussian distribution is an attractor for sums of i.i.d. variables under additive normalization, EDMs are fixed points of a renormalization involving both addition and scale transformation of the mean.
Specifically, if we let
\begin{equation}\label{sm:eq:random_sum}
    Y_N = \frac{1}{a_N}\sum_{k=1}^N X_k
\end{equation}
where each $X_k>0$ is a positive random variable and $a_N$ is a normalization factor that preserves the variance–mean structure.
Then, as $N \rightarrow \infty$, the distribution of $Y_N$ approaches a Tweedie distribution. 

Within the context of our framework, the relative abundance of each component is the sum of $N$ binary variates~(Supplementary Note~\ref{sm:sampling}), i.e.
\begin{equation}
    n_i = \sum_{s=1}^N X_i^{(s)}
\end{equation}
Again letting relative counts be $\nu_i = n_i / N$, then $n_i$ is exactly a variable as in Eq.~\ref{sm:eq:random_sum} with $a_N = N$.
Supposing independent $X_k$ with mean $\mu$ and with asymptotical variance $\operatorname{Var}\!\left[X\right] \sim c\mu^b$, then the mean and variance of the sum are
\begin{equation}
    \operatorname{E}\!\left[Y_N\right] = N \mu, \qquad \operatorname{Var}\!\left[Y_N\right] = N c \mu^b
\end{equation}
Rewriting variance in terms of the new mean $\operatorname{E}\!\left[Y_N\right] := \mu_N$ leads to
\begin{equation}\label{sm:eq:var_sum}
    \operatorname{Var}\!\left[Y_N\right] = c \mu_N^b N^{1-b}.
\end{equation}
In general, the renormalization factor $a_N$ must be chosen so that the extra factor $N^{1-b}$ cancels.
It is straightforward to see that $a_N = N$ respect this condition, thus relative counts $\nu_i$ are random variables with a limiting distribution in the Tweedie family with Tweedie exponent $b$.

\begin{figure}[htbp]
  \centering

  \begin{subfigure}[t]{0.48\textwidth}
    \centering
    \includegraphics[width=\linewidth]{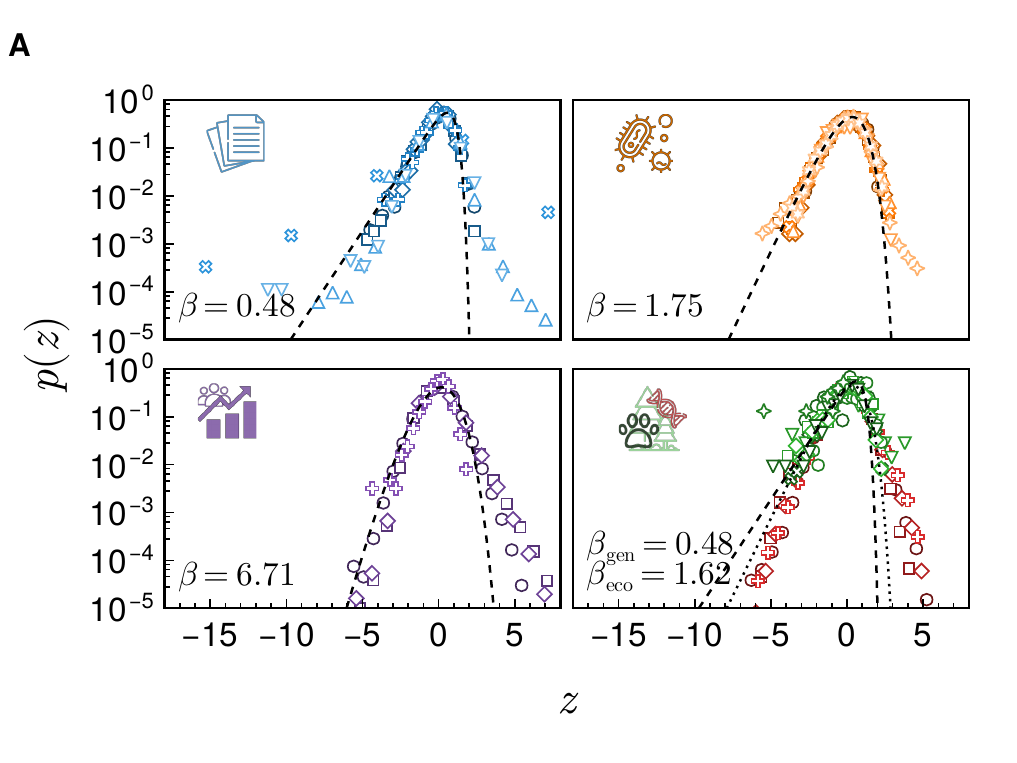}
    \phantomcaption
    \label{sm:fig:afd_a}
  \end{subfigure}\hfill
  \begin{subfigure}[t]{0.48\textwidth}
    \centering
    \includegraphics[width=\linewidth]{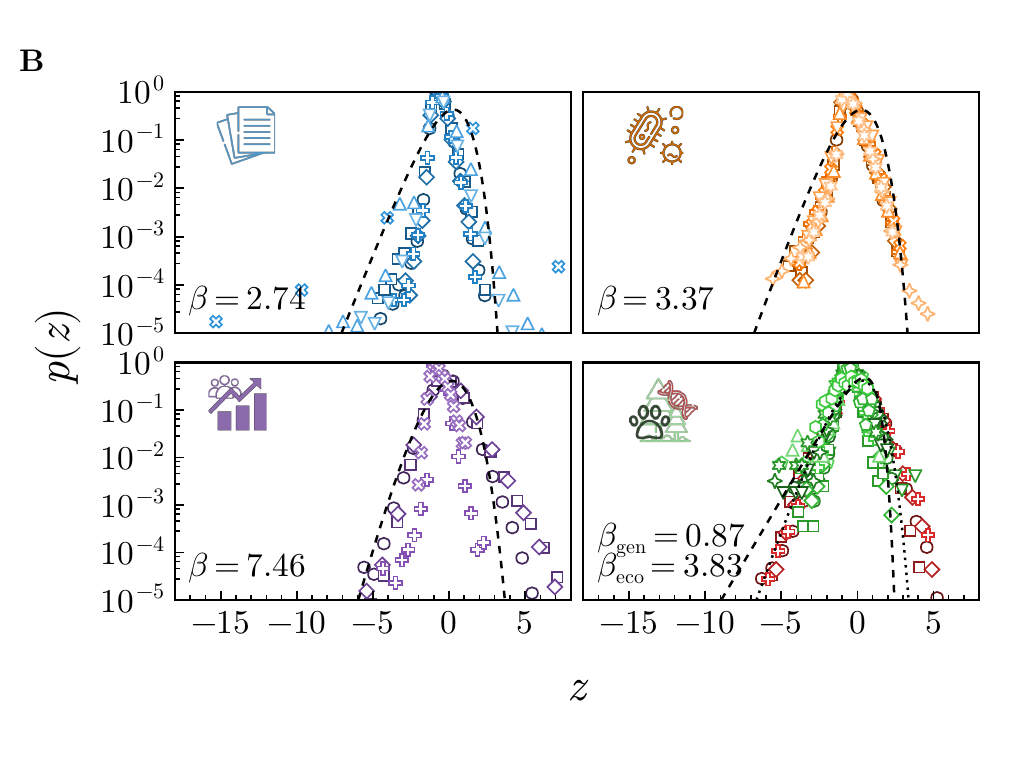}
    \phantomcaption
    \label{sm:fig:afd_b}
  \end{subfigure}

  \vspace{-1.5\baselineskip}

  \begin{subfigure}[t]{0.48\textwidth}
    \centering
    \includegraphics[width=\linewidth]{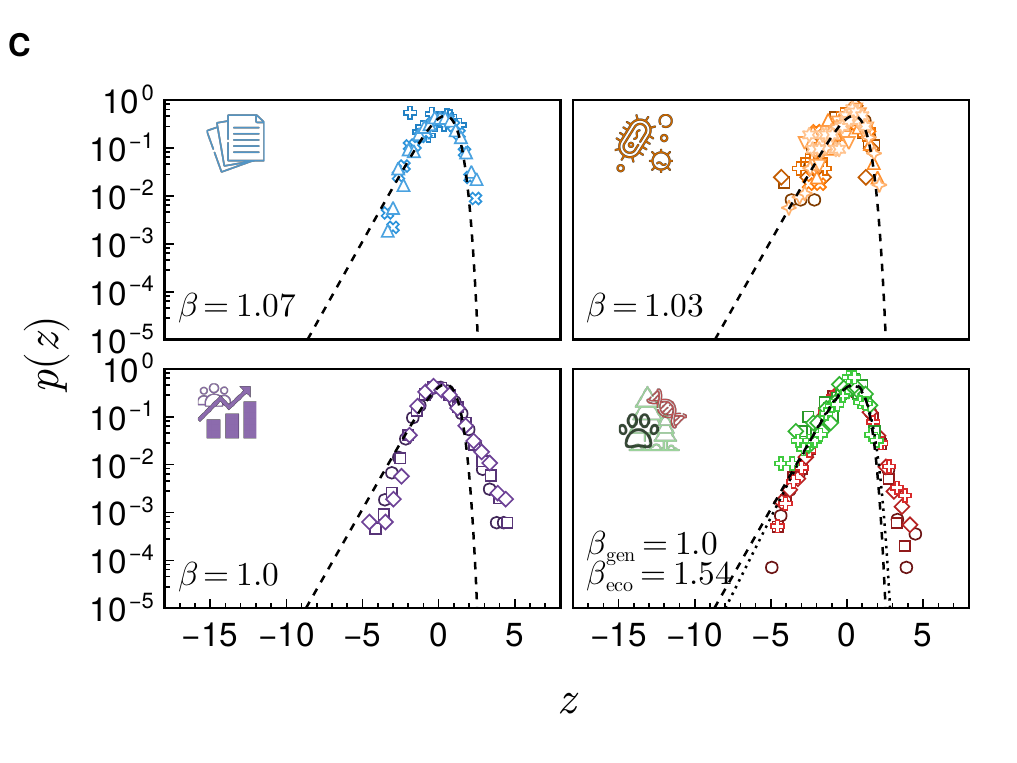}
    \phantomcaption
    \label{sm:fig:afd_c}
  \end{subfigure}\hfill
  \begin{subfigure}[t]{0.48\textwidth}
    \centering
    \includegraphics[width=\linewidth]{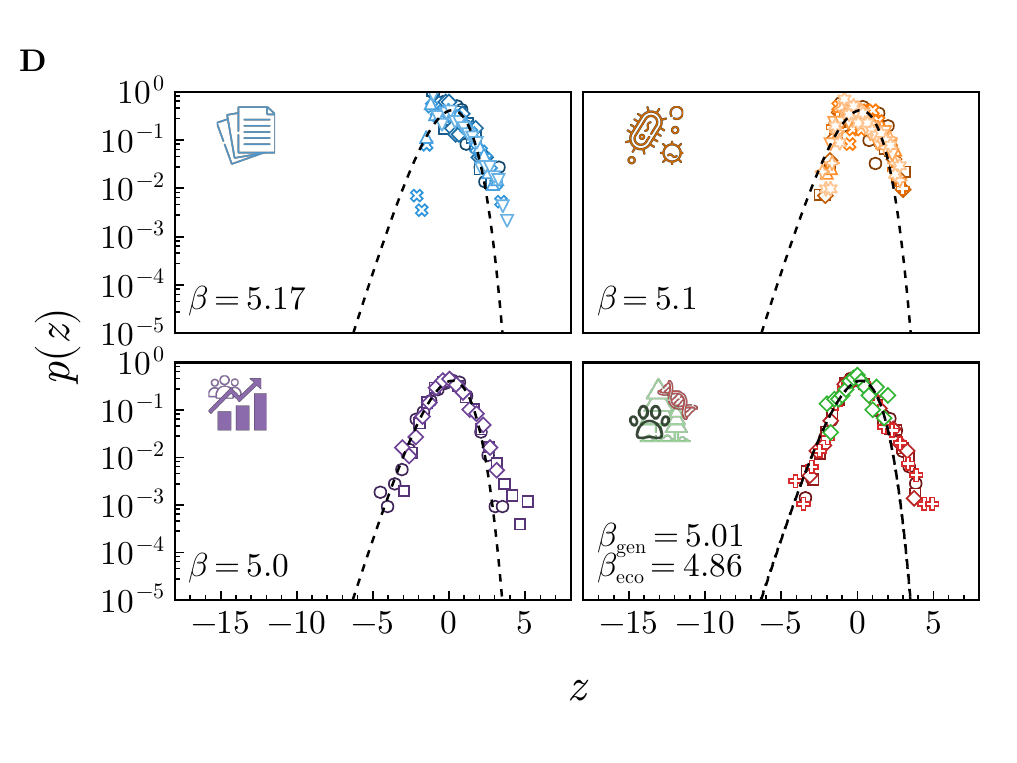}
    \phantomcaption
    \label{sm:fig:afd_d}
  \end{subfigure}

  \vspace{-2.3\baselineskip}
  \caption{
    \textbf{Abundance fluctuation distribution at different levels of occupancy.}
    (\textbf{A}) When restricting to components appearing in $99\%$ of samples ($99\%$ occupancy level), Taylor's exponent is pushed towards $b \rightarrow 2$, leading to a fluctuation distribution that is approximated by a Gamma distribution (dashed black and dotted gray lines) with inverse shape parameter $\beta= \sigma^2 / \mu^2$.
    The figure shows the distribution of the rescaled variable $z=(y - m) / s$, where $m=\operatorname{E}\!\left[\log\nu\right]$ and $s = \sqrt{\operatorname{Var}\!\left[\log\nu\right]}$; see~\cite{grilli2020macroecological}.
    Notice, however, that high occupancy does not necessarily imply $b \approx 2$: apart from the microbial case, the Gamma distribution gives only a rough approximation of the AFD for the other datasets.
    (\textbf{B}) Lower occupancy ($50\%$) corresponds to $1<b<2$, where the Gamma approximation breaks down.
    (\textbf{C}, \textbf{D}) The Gamma approximation is improved when restricting to components with $\mu_i \gg \Omega_i^{-1}$.
    Figures show the AFD obtained for components with respectively $\Omega_i \simeq 1$ and $\Omega_i \simeq 5$ at $50\%$ occupancy level.
  }\label{sm:fig:afd}
\end{figure}

Relating back to Taylor's law, we showed previously how common components exhibit $b \rightarrow 2$.
As a result, the associated Tweedie distribution of component abundance fluctuations is the Gamma distribution.
It is important to realize that the Gamma distribution has previously been put forward as a proper candidate for the AFD in microbial systems (see, e.g., \cite{grilli2020macroecological,camacho-mateu2024sparse,camacho-mateu2024nonequilibrium} and Fig.~\ref{sm:fig:afd}).
As such, these so-called \emph{macroecological laws} therefore also emerge naturally from the proposed framework.

Moreover, the Gamma AFD can be tested more directly by selecting components according to their quadratic coefficients rather than by occupancy alone.
The reason is that the crossover to the quadratic regime occurs when $\mu_i > \Omega_i^{-1}$.
Thus, instead of imposing an occupancy threshold, one can retain only those components for which the empirical mean lies beyond this crossover scale.
For these components, the linear sampling term becomes subdominant and the mean--variance relation reduces to
\begin{equation}
  \operatorname{Var}\!\left[n_i\right] \simeq \Omega_i \operatorname{E}\!\left[n_i\right]^2 .
\end{equation}
Since $\Omega_i$ corresponds to the squared coefficient of variation, this is precisely the mean--variance relation associated with Gamma-distributed abundance fluctuations.
Therefore, the same proposed framework not only recovers the emergence of the Gamma AFD, but also provides a coefficient-based criterion for identifying the regime in which this macroecological law is expected to hold.

\subsection{Taylor's law from Binomial sampling}\label{sm:binomial}
Taylor's law (and the other scaling laws below) may also be derived considering different sampling schemes.
For example, one may assume that counts are generated by a Binomial process instead of a Poisson process, so that $n_{i;k} \sim \operatorname{Binomial}(N_k, p_{i;k})$.
In this case, the observer effectively chooses the sample size $N_k$ instead of having it be a random variate reflecting systems that have a fixed (maximum) number of possible observations, for example due to technical limitations or sampling costs.

Consider this sampling scheme with fixed $N_k = N$ for simplicity (the general case is equal to that considered in~Supplementary Note~\ref{sm:samplingheterogeneity}), and success probability $p$, which is a random variable so that $n_i \mid p_i \sim \mathrm{Bin}(N,p_i)$.
Here, $p_i$ plays the role of the latent variable.

Conditional moments are given by
\begin{equation}
  \operatorname{E}\!\left[n_i\mid N ,\, p_i\right] = N p_i, \qquad \operatorname{Var}\!\left[n_i\mid N ,\, p_i\right] = N p_i (1-p_i).
\end{equation}
Using the law of total variance and the conditional binomial moments, one obtains
\begin{subequations}
  \begin{align}
    \operatorname{Var}\!\left[n_i\right] &= N^2 \operatorname{Var}\!\left[p_i\right] + N \operatorname{E}\!\left[p_i(1-p_i)\right] \\
              &= N^2 \operatorname{Var}\!\left[p_i\right] + N \operatorname{E}\!\left[p_i\right] - N \operatorname{E}\!\left[p_i^2\right] \\
              &= N^2 \operatorname{Var}\!\left[p_i\right] + \operatorname{E}\!\left[n_i\right] - \frac{\operatorname{E}\!\left[p_i^2\right]}{\operatorname{E}\!\left[p_i\right]}\operatorname{E}\!\left[n_i\right] \\
              &= \operatorname{E}\!\left[n_i\right]^2 \frac{\operatorname{Var}\!\left[p_i\right]}{E[p_i]^2} + \operatorname{E}\!\left[n_i\right]
                \left(1 - \frac{\operatorname{Var}\!\left[p_i\right] - \operatorname{E}\!\left[p_i\right]^2}{E[p_i]} \right)
  \end{align}
\end{subequations}
which is in the same form as Eq.~(2) in the main text found for the Poisson sampling scheme but with different coefficients.
Similar arguments may be considered that will lead to the other two scaling laws, Zipf's and Heaps' law, under a binomial sampling scheme.

\section{Additional structures}\label{sm:additional_struct}
Altering the sampling procedure may result in violation of De Finetti's assumptions.
For example, considering a non-memoryless scheme such as sequential reading of components, or a biased scheme for which new observations depends on the number of old observations, will inevitably add internal correlations which would alter Eq.~(2) in the main text.
Thus, these additional structures may be expected to alter Taylor's law allowing to reflect the scaling behavior found empirically in more exotic systems~\cite{eisler2008fluctuation}.
For example, if a sequence of observation of component $i$ is composed of correlated Bernoulli variables, then one would find
\begin{equation}
    \operatorname{Var}\!\left[n_i\right] = \sum_t \operatorname{Var}\!\left[X_i^{(t)}\right] + 2\sum_{t<s}\operatorname{cov}\left[X_i^{(t)} X_i^{(s)}\right].
\end{equation}
This adds complications to Eq.~\ref{sm:eq:var}, possibly creating systematic deviations from a convergence to $b=2$.
A systematic treatment of correlated or history-dependent sampling schemes lies beyond the scope of the present work and is left for future investigation.

\section{Zipf's law as a consequence of heavy-tailed component abundances}\label{sm:zipf}
Zipf's law follows immediately from heavy-tailed component abundances.
To see why, let us consider a positive random variable $X$ distributed according to a power law,
\begin{equation}
  p(x) \propto x^{-\alpha}, \qquad x \ge x_{\min}, \quad \alpha > 1,
\end{equation}
The complementary cumulative distribution function (CCDF) is defined as
\begin{equation}
  \bar{F}(x) = \Pr(X \ge x) = \int_x^{\infty} p(t)\,dt \propto x^{-(\alpha-1)}
\end{equation}
Letting $\beta = \alpha - 1$, the CCDF can be written as $\bar{F}(x) \propto x^{-\beta}$.
Now consider a sample of size $N$ drawn independently from this distribution and sorted in decreasing order,
\begin{equation}
  x_{(1)} \ge x_{(2)} \ge \cdots \ge x_{(N)},
\end{equation}
where $x_{(r)}$ denotes the value of the component with rank $r$. By definition, the rank $r$ corresponds to the number of observations with value greater than or equal to $x_{(r)}$. In expectation, this number is given by
\begin{equation}
  r \approx N \, \bar F\bigl(x_{(r)}\bigr).
\end{equation}
Using the power-law form of the CCDF, we obtain $r \propto x_{(r)}^{-\beta}$, and inverting this relation yields
\begin{equation}
  x_{(r)} \propto r^{-1/\beta} \equiv r^{-\zeta},
\end{equation}
which is Zipf's law with exponent
\begin{equation}
  \zeta = \frac{1}{\beta} = \frac{1}{\alpha - 1}.
\end{equation}
As such, Zipf's law follows directly from the power law form of the CCDF, as stated earlier.
This relation shows that Zipf's law is simply the rank representation of an underlying power law distribution.
In our case, these power laws are the heavy-tailed CADs as defined above and in the main text.

\section{Vocabulary growth and Heaps' law in complex component systems}\label{sm:heaps}
Heaps' law regards scaling of the so-called \emph{vocabulary size}.
That is, it relates the vocabulary size $V(N)$ with the total (accumulated) sample size $N$ as $V(N) \propto N^{\eta}$~\cite{heaps1978information,herdan1958relation,vanleijenhorst2005formal}, where $0 \leq \eta \leq 1$.
To show that this scaling emerges from the latent distribution, we start by writing the probability of observing a specific component in an observation as $\Pr[n_{ik} > 0] = 1 - \Pr[n_{ik} = 0]$.
We are interested in the expected vocabulary size across observations $\operatorname{E}\!\left[V(N)\right] = \operatorname{E}_k\!\left[V(N)\right]$, which for a Poisson sampling scheme reads
\begin{equation}
  \operatorname{E}\!\left[V(N)\right] = \sum_{i=1}^S \left( 1 - e^{-N\theta_i} \right)
\end{equation}
where we have omitted the explicit observation index $k$ under the assumption that the latent distribution does not change drastically between samples.
That this is the case can be seen by analysis of CAD exponents $\gamma$ across observations (Fig.~3 in the main text).
These distributions are relatively narrow and appear to even be stable across environments within a specific domain, which makes the assumption above very reasonable.
Note that this does \emph{not} mean that the latent intensities $\theta_{ik}$ do not change between distribution.
Rather, it is the distribution $\theta_{ik} \sim p(\theta)$ that does not change drastically --- i.e., exponents of the heavy-tailed distribution $p(\theta)$ are relatively stable.

Now, as vocabulary growth must, by definition, saturate at some point due to systems being inherently finite, let us formally introduce a characteristic scale that reflects this.
That is, consider that the system is finite and contains $S < \infty$ distinct components (i.e., the maximum vocabulary size is $S$), and at the time of an observation each component appears exactly $m_i$ times, with the total number of components $M = \sum_i m_i < \infty$.
In such a system, there must always be a specific component for which its latent intensity is smallest.
In other words, there is a rarest component with intensity $\theta_{\min} = \min_i \theta_i$.
An observation is thus expected to saturate when it is of length $\varphi \sim 1 / \theta_{\min}$.
That is, for $N \ll \varphi$, the rarest component(s) remain unseen, whereas for $N \gg \varphi$ all components are likely to have been observed.
We may write $\theta_{\min} \sim m_{\min} / (M \operatorname{E}\!\left[m_i\right])$ and thus the characteristic scale can be written to be
\begin{equation}
  \varphi \sim \frac{M \operatorname{E}\!\left[m_i\right]}{m_{\min}}
\end{equation}
Using this scale, we can define a generic counting function $F(\theta) = \# \{i : \theta_i \geq \theta\}$ which simply counts the number of components that has intensity $\theta$.
This counting function is useful as it relates to the characteristic scale as the number of saturated components (i.e., those that have probability approximately $1$ of being observed after $N$ samples) are given by $F(1/N)$.
That is, when $N \gg \varphi$, we have that $F(1/\varphi) = F(\theta_{\min}) = M$, whereas for the small and intermediate regimes where $N \ll \varphi$ and $N \sim \varphi$ respectively we find Heaps' law instead.
To see this, we make the general assumption that the counting function reads
\begin{equation}
  F(\theta) =
  \begin{cases}
    M, &\theta \leq \theta_{\min},\\
    G(\theta) &\theta > \theta_{\min}
  \end{cases}
\end{equation}
with
\begin{equation}
  G(\theta) \sim L(\theta) \theta^{-\eta}  
\end{equation}
with $L(\theta)$ a slowly varying function.
Notice that this function is proportional to the CDF of the variable $\theta$.

Let us first tackle the regimes where $N \ll \varphi$ and $N \gg \varphi$.
In the former, we expect that for almost all components we have $\Pr[n_i=0] \approx 1$, which is the case for $N\theta_i \ll 1$.
Therefore, the exponent can be approximated as $e^{-N\theta_i} \approx 1 - N\theta_i$, and we get
\begin{equation}
  \operatorname{E}\!\left[V(N)\right] \approx \sum_i \left[ 1 - (1 - N\theta_i) \right] \propto N  
\end{equation}
and linear growth follows.
In the latter regime where all components are saturated, we get that $\Pr[n_i=0]\rightarrow 0$ and it is straightforward to see that the vocabulary size saturates as
\begin{equation}
  \operatorname{E}\!\left[V(N)\right] \rightarrow S
\end{equation}
However, in practice, this saturation is almost never observed due to $N \ll \varphi$.
For example, there exists no book that contains every word in a language at least once.
Therefore, from the observer's perspective the maximum admissible vocabulary size $S$ is practically infinite.

In contrast, when $N \sim \varphi$ only some components have saturated while others have not.
Using the counting function introduced above, the expected vocabulary size in this regime reads
\begin{equation}
  \operatorname{E}\!\left[V(N)\right] = F(1/N) + N \Theta(1/N)
\end{equation}
where
\begin{equation}
  \Theta(1/N) = \sum_{\substack{j \\ \theta_j \ll 1/N}} \theta_j
\end{equation}
with the sum going over intensities that are (significantly) smaller than the resolution of the observation $1/N$.
For the first term, we use the definition of the counting function and note that for these components $\theta_i > \theta_{\min}$ and thus we have $F(1/N) \sim N^\eta$.
We approximate the sum $\Theta$ by an integral noting that
\begin{equation}
  \frac{dF(\theta)}{d\theta} = -c \gamma \theta^{-\eta-1}
\end{equation}
with $dL(\theta)/d\theta \approx c$ a constant equal to the derivative of the slowly varying function as $\theta \rightarrow 0$ (for $\theta < \theta_{\min}$).
Then, the sum is approximated as
\begin{equation}
  N \Theta
  \sim N \int_0^{1/N} \theta dF(\theta)
  = \frac{c\eta}{1-\eta} N^{\eta} \propto N^{\eta}.
\end{equation}
Thus, as both terms scale with $N^\eta$, Heaps' law with exponent $\eta$ emerges in the intermediate regime where the sampling depth is enough to overcome the linear discovery phase yet not sufficient to fully resolve the rarest components.
Naturally, this is the regime wherein most observations of real systems are made.
When the characteristic scale of the system is so large that reasonable observations of length $N$ are still in the intermediate regime, Heaps' law appears to suggest that discovery of novel components will carry on seemingly forever.
However, it is important to realize that in any finite system with $M<\infty$ one will eventually observe convergence of the vocabulary size to the true number of distinct components.

\paragraph{Bounds on vocabulary growth}
However, do note that a fundamental bound must be taken into account. 
Since a sample of size $N$ can contain at most $N$ distinct components, the growth of the vocabulary cannot exceed linear scaling. 
This imposes a strict upper limit on the Heaps' exponent,
\begin{equation}
    \eta = \min(\gamma - 1,\, 1).
\end{equation}
This bound reflects a basic constraint of counting statistics and holds independently of the specific form of the underlying abundance distribution.

\clearpage
\section*{Tables}
\subsection*{Data sources and filtering}

\begin{table*}[htbp]
  \centering
  \small
  \caption{
    Overview of the data used in this study including their original source.
    We include their field of research, on what criterion they were partitioned, and what constitutes a single sample.
    Colors are indicative for their field as in~Figs.~2 and~3 in the main text.
    When no partition is reported, the dataset is not subdivided and used as is.
  }
  \resizebox{\textwidth}{!}{
  \begin{NiceTabular}{l l l l c c}
    \CodeBefore
    \rowcolor{gray!50!white}{1}
    \rowcolor{CornflowerBlue!15}{2,3,4}
    \rowcolor{YellowOrange!15}{5}
    \rowcolor{Orchid!15}{6,7,8}
    \rowcolor{Green!15}{9,10}
    \rowcolor{Red!15}{11}
    \Body
    \toprule\RowStyle{\bfseries}
    dataset & field & partitioned on & sample & \#samples & source \\
    \midrule
    arXiv articles & linguistics & category & article & 846 & \cite{arxiv} \\
    Project Gutenberg & linguistics & language & book & 213 & \cite{project_gutenberg} \\
    RFC Series & linguistics & -- & document & 158 & \cite{rfc_archive} \\
    \hdottedline
    EBI Metagenomics & microbiology & environment & sample & 2005 & \cite{mitchell2018ebi,grilli2020macroecological} \\
    \hdottedline
    Gowalla project & human activity & -- & days & 301 & \cite{cho2011friendship} \\
    LEGO catalog & human activity & -- & LEGO set & 118 & \cite{rebrickable} \\
    Stock volumes & financial & financial markets & daily volumes & 755 & \cite{yahoo_finance} \\
    \hdottedline
    BCI trees & ecology & -- & quadrat & 256 & \cite{condit2012barro} \\
    BioTIME & ecology & project & sample & 264 & \cite{dornelas2018biotime} \\
    \hdottedline
    Genotype-Tissue Expression (GTEx) & biology & tissue type & sample & 800 & \cite{gtex2020atlas} \\
    \bottomrule
  \end{NiceTabular}
  }
  \label{sm:tab:data}
\end{table*}

\begin{table*}[htbp]
  \centering
  \small
  \caption{
    Filtering criteria applied to each dataset.
    All partitions and samples that did not meet the criteria were omitted from the analysis.
    Parameters include the minimum total counts per sample, the minimum number of samples per class, the minimum number of distinct components per partition, and the minimum number of distinct components per sample.
    These constraints were chosen to omit samples with very few components, but still to retain a significant portion of the original dataset.
    We verified empirically that results do not change drastically when these parameters change slightly.
  }
    \resizebox{\textwidth}{!}{
    \begin{NiceTabular}{l c c c c}
      \CodeBefore
      \rowcolor{gray!50!white}{1}
      \rowcolor{CornflowerBlue!15}{2,3,4}
      \rowcolor{YellowOrange!15}{5}
      \rowcolor{Orchid!15}{6,7,8}
      \rowcolor{Green!15}{9,10}
      \rowcolor{Red!15}{11}
      \Body
      \toprule\RowStyle{\bfseries}
      dataset & min. counts & min. samples & distinct components  & min. components \\
      \midrule
      arXiv & $4000$ & 30 & 1000 & 500 \\
      Project Gutenberg & $10^5$ & 30 & 1000 & 500 \\
      RFC archive & $10^4$ & 30 & 1000 & 500 \\
      \hdottedline
      EBI Metagenomics & $10^4$ & 30 & 500 & 200 \\
      \hdottedline
      Gowalla project & $10^4$ & 30 & 500 & 200 \\
      LEGO catalog & $10^4$ & 30 & 500 & 100 \\
      Financial markets & $10^4$ & 30 & 500 & 200 \\
      \hdottedline
      BCI trees & $5000$ & 30 & 200 & 100 \\
      BioTIME database & 5000 & 30 & 200 & 100 \\
      \hdottedline
      GTEx project & $10^8$ & 30 & 1000 & 500 \\
      \bottomrule
    \end{NiceTabular}
    }
    \label{sm:tab:filter}
\end{table*}

\clearpage
\subsection*{Taylor's law model comparison}

\begin{table*}[htbp]
  \centering
  \caption{
    Comparison between the quadratic and power-law forms of Taylor's law.
    Fits are performed separately on groups of components with similar empirical quadratic coefficients $\hat{\Omega}_i$ and containing at least $20$ components.
    For each dataset, we report the model selected most often according to the lowest AIC, the fraction of component groups selecting the power-law model, and the mean AIC and log-RMSE values across groups.
    }
  \scriptsize
  \setlength{\tabcolsep}{3pt}
  \renewcommand{\arraystretch}{1.05}

  \begin{NiceTabular}{l l r r r r r}
    \CodeBefore
    \rowcolor{gray!50!white}{1}
    \rowcolor{CornflowerBlue!15}{2-8}
    \rowcolor{YellowOrange!15}{9-17}
    \rowcolor{Orchid!15}{18-22}
    \rowcolor{Green!15}{23}
    \rowcolor{Red!15}{24-27}
    \Body
    \toprule\RowStyle{\bfseries}
    dataset & best model & $f_{\mathrm{power}}$ & $\operatorname{E}\!\left[\mathrm{AIC}_{\mathrm{quad}}\right]$ & $\operatorname{E}\!\left[\mathrm{AIC}_{\mathrm{power}}\right]$ & $\operatorname{E}\!\left[\log\mathrm{RMSE}_{\mathrm{quad}}\right]$ & $\operatorname{E}\!\left[\log\mathrm{RMSE}_{\mathrm{power}}\right]$ \\
    \midrule
    arx-astro-ph & quadratic & 0.373 & 75.074 & 78.366 & 0.429 & 0.426 \\
    arx-math & quadratic & 0.312 & 56.184 & 58.790 & 0.433 & 0.433 \\
    arx-physics & quadratic & 0.420 & 93.877 & 93.989 & 0.494 & 0.482 \\
    arx-q-bio & quadratic & 0.321 & 87.632 & 88.571 & 0.478 & 0.469 \\
    guten-en & quadratic & 0.435 & 552.393 & 561.835 & 0.609 & 0.601 \\
    guten-it & quadratic & 0.348 & 858.108 & 897.690 & 0.511 & 0.513 \\
    rfc-rfc & quadratic & 0.400 & 129.717 & 131.810 & 0.525 & 0.522 \\
    \hdottedline
    otu-AQUA1 & quadratic & 0.170 & 149.446 & 164.162 & 0.530 & 0.585 \\
    otu-GUT1 & quadratic & 0.250 & 118.302 & 122.708 & 0.580 & 0.586 \\
    otu-GUT2 & quadratic & 0.123 & 180.279 & 193.193 & 0.594 & 0.620 \\
    otu-LAKE & quadratic & 0.333 & 100.596 & 103.628 & 0.642 & 0.638 \\
    otu-ORAL1 & quadratic & 0.130 & 52.666 & 58.947 & 0.476 & 0.503 \\
    otu-RIVER & quadratic & 0.138 & 147.932 & 158.233 & 0.565 & 0.586 \\
    otu-SEA & quadratic & 0.000 & 28.826 & 30.261 & 0.431 & 0.425 \\
    otu-SLUDGE & quadratic & 0.108 & 230.969 & 245.530 & 0.606 & 0.628 \\
    otu-SOIL & quadratic & 0.088 & 46.085 & 51.166 & 0.555 & 0.586 \\
    \hdottedline
    gowalla-gowalla & quadratic & 0.492 & -54.990 & -52.712 & 0.212 & 0.204 \\
    LEGO & quadratic & 0.391 & 23.746 & 23.765 & 0.375 & 0.361 \\
    finance-euronext & quadratic & 0.083 & 46.324 & 47.460 & 0.627 & 0.616 \\
    finance-nasdaq & quadratic & 0.244 & 148.269 & 149.018 & 0.879 & 0.869 \\
    finance-nyse & quadratic & 0.143 & 87.162 & 88.358 & 0.570 & 0.564 \\
    \hdottedline
    biotime-129 & quadratic & 0.000 & 26.129 & 31.550 & 0.409 & 0.440 \\
    \hdottedline
    gtex-BRAIN & power & 0.867 & 938.652 & 903.531 & 0.561 & 0.544 \\
    gtex-HEART & power & 0.902 & 665.509 & 619.544 & 0.480 & 0.456 \\
    gtex-LIVER & power & 0.914 & 848.548 & 770.977 & 0.570 & 0.526 \\
    gtex-PANCREAS & power & 0.761 & 967.333 & 923.223 & 0.627 & 0.601 \\
    \bottomrule
  \end{NiceTabular}
  \label{sm:tab:taylor_model_comparison}
\end{table*}

\clearpage
\subsection*{Results of fitting distributions}

\begin{table*}[htbp]
  \centering
  \caption{
    For each (partitioned) dataset we list the fraction of samples in which a heavy-tailed distribution was found best to describe the data.
    Additionally, we list the model that was most likely to describe the data as the one with the lowest Akaike information criterion (see~Methods).
    We also list mean and variance of the computed $p$ values across all samples (thus also including those for which a heavy-tailed distribution was rejected).
    Note that only counts from the gtex-PANCREAS class almost never exhibit heavy tails.
    Additionally note the large variance in LEGO on the exponent, which likely is because LEGO sets typically contain relatively few distinct pieces and thus tails span less than a decade or two.
    Finally, mean $\operatorname{E}\!\left[\gamma\right]$ and variance $\operatorname{Var}\!\left[\gamma\right]$ of the power law exponent $\gamma$ for each dataset.
    The exponent is the exponent of the distribution of the most likely model.
    Note that for most datasets the exponent values do not differ significantly between datasets as indicated by low values of $\operatorname{Var}\!\left[\gamma\right]$, which indicates that system-specific processes must be at play.
  }
  \small
  \setlength{\tabcolsep}{5pt}
  \renewcommand{\arraystretch}{1.05}

  \begin{NiceTabular}{l c l c c c c}
    \CodeBefore
    \rowcolor{gray!50!white}{1}
    \rowcolor{CornflowerBlue!15}{2-8}
    \rowcolor{YellowOrange!15}{9-17}
    \rowcolor{Orchid!15}{18-22}
    \rowcolor{Green!15}{23-28}
    \rowcolor{Red!15}{29-32}
    \Body
    \toprule
    \textbf{dataset} & $f_{\textrm{ht}}$ & \textbf{most likely model} & $\operatorname{E}\!\left[p\right]$ & $\operatorname{Var}\!\left[p\right]$ & $\operatorname{E}\!\left[\gamma\right]$ & $\operatorname{Var}\!\left[\gamma\right]$ \\
    \midrule
    arxiv-physics & 0.84 & ParetoIV & 0.46 & 0.106 & 2.34 & 0.037\\
    arxiv-math & 0.8 & TemperedPareto & 0.53 & 0.109 & 2.02 & 0.03\\
    arxiv-astro-ph & 0.9 & ParetoIV & 0.6 & 0.095 & 2.24 & 0.029\\
    arxiv-q-bio & 0.88 & ParetoIV & 0.49 & 0.1 & 2.35 & 0.032\\
    gutenberg-en & 0.86 & ParetoIV & 0.49 & 0.097 & 2.03 & 0.011\\
    gutenberg-it & 0.86 & ParetoIV & 0.5 & 0.113 & 2.02 & 0.009\\
    rfc-rfc & 0.6 & TemperedPareto & 0.32 & 0.11 & 1.99 & 0.047\\
    \hdottedline
    otu-AQUA1 & 0.92 & TemperedPareto & 0.57 & 0.067 & 1.66 & 0.007\\
    otu-GUT1 & 0.8 & TemperedPareto & 0.56 & 0.124 & 1.64 & 0.024\\
    otu-GUT2 & 0.76 & TemperedPareto & 0.41 & 0.108 & 1.62 & 0.007\\
    otu-LAKE & 0.8 & TemperedPareto & 0.48 & 0.127 & 1.56 & 0.011\\
    otu-ORAL1 & 0.86 & TemperedPareto & 0.48 & 0.094 & 1.58 & 0.024\\
    otu-RIVER & 0.76 & TemperedPareto & 0.46 & 0.119 & 1.73 & 0.024\\
    otu-SEA & 0.94 & TemperedPareto & 0.63 & 0.066 & 1.4 & 0.004\\
    otu-SLUDGE & 0.42 & TemperedPareto & 0.24 & 0.092 & 1.77 & 0.014\\
    otu-SOIL & 0.86 & ParetoIV & 0.44 & 0.099 & 1.84 & 0.022\\
    \hdottedline
    gowalla-gowalla & 0.3 & TemperedPareto & 0.26 & 0.18 & 2.06 & 0.008\\
    LEGO & 0.54 & TemperedPareto & 0.26 & 0.077 & 2.37 & 1.606 \\
    finance-euronext & 0.9 & ParetoIV & 0.47 & 0.084 & 1.7 & 0.004\\
    finance-nasdaq & 0.96 & ParetoIV & 0.59 & 0.078 & 1.94 & 0.002\\
    finance-nyse & 0.86 & ParetoIV & 0.41 & 0.077 & 2.33 & 0.011\\
    \hdottedline
    bcitrees-BCI & 0.46 & ParetoIV & 0.45 & 0.109 & 2.16 & 0.278\\
    biotime-129 & 0.6 & ParetoIV & 0.35 & 0.084 & 1.82 & 0.073\\
    biotime-200 & 0.73 & ParetoIV & 0.6 & 0.118 & 1.69 & 0.053\\
    biotime-582 & 0.4 & ParetoIV & 0.38 & 0.105 & 1.45 & 0.018\\
    biotime-604 & 0.85 & ParetoIV & 0.5 & 0.098 & 2.24 & 0.47\\
    biotime-671 & 0.74 & ParetoIV & 0.47 & 0.116 & 1.97 & 0.289\\
    gtex-BRAIN & 0.68 & ParetoI & 0.32 & 0.091 & 2.45 & 0.015\\
    gtex-HEART & 0.32 & ParetoI & 0.1 & 0.02 & 2.25 & 0.003\\
    gtex-LIVER & 0.42 & ParetoI & 0.15 & 0.033 & 2.16 & 0.005\\
    gtex-PANCREAS & 0.08 & ParetoI & 0.04 & 0.025 & 2.23 & 0.025\\
    \bottomrule
  \end{NiceTabular}  
  \label{sm:tab:fitting}
\end{table*}

\clearpage
\subsection*{Candidate distributions}

\begin{table*}[htbp]
  \centering
  \caption{
    Proposed distributions with model index $w$ (see~Methods) to fit component abundance distributions (CADs), including their probability density function $f(\nu)$ and survival function $\bar{F}(\nu) = 1 - F(\nu)$ for $\nu \ge \varepsilon$, and the set of parameters to be inferred.
    For nearly all datasets that we analyzed a member of the Pareto family of distributions was often found to best describe the tail of abundance distributions.    
    Note the differences in domain for distributions of the Pareto family and the other distributions.
    \textsuperscript{a}$\gamma_{\textrm{inc}}$ and $\Gamma$ are the incomplete gamma function and the gamma function, respectively.
    \textsuperscript{b}$\Phi$ is cumulative distribution function of the standard normal distribution.
  }
  \setlength{\extrarowheight}{3pt}
  \begin{small}
    \resizebox{\textwidth}{!}{
    \begin{NiceTabular}{l l l l l}
      \CodeBefore
      \rowcolor{gray!50!white}{1}
      \Body
      \toprule\RowStyle{\bfseries}
      {\sffamily $w$} & distribution & pdf $f(\nu)$ & survival $\bar{F}(\nu)$ & parameters \\
      \midrule
      1 & Pareto I 
        & $\alpha \varepsilon^\alpha \nu^{-(\alpha+1)}$ 
        & $(\nu / \varepsilon)^{-\alpha}$ 
        & $\alpha, \varepsilon > 0$ \\

      2 & Pareto IV 
        & $\frac{\alpha}{\beta \omega} \left(\frac{\nu - \varepsilon}{\omega}\right)^{1/\beta - 1} \left[1 + \left(\frac{\nu - \varepsilon}{\omega}\right)^{1/\beta}\right]^{-(\alpha+1)}$
        & $\left[ 1 + \left(\frac{\nu - \varepsilon}{\omega}\right)^{1/\beta} \right]^{-\alpha}$ 
        & $\alpha, \beta, \omega ,\varepsilon > 0$ \\

      3 & tempered Pareto 
        & $C\, \nu^{-(\alpha+1)} e^{-\beta \nu}$ 
        & $\varepsilon^d e^{\beta \varepsilon} \nu^{-\alpha} e^{-\beta \nu}$ 
        & $\alpha,\beta,\varepsilon > 0$ \\

      4 & Gamma\textsuperscript{a}
        & $\frac{1}{\Gamma(\alpha)\,\omega^\alpha} \nu^{\alpha-1} e^{-\nu/\omega}$ 
        & $1 - \gamma_{\textrm{inc}}(\alpha, \nu/\omega) / \Gamma(\alpha)$ 
        & $\alpha,\omega > 0$ \\

      5 & log-normal\textsuperscript{b}
        & $\frac{1}{\nu \sigma \sqrt{2\pi}} \exp\!\left[-\frac{(\log \nu - \mu)^2}{2\sigma^2}\right]$ 
        & $1 - \Phi\left((\log \nu - \mu)/\sigma \right)$ 
        & $\mu \in \mathbb{R}, \sigma > 0$ \\

      6 & Weibull 
        & $\frac{k}{\omega} \left(\frac{\nu}{\omega}\right)^{k-1} e^{-(\nu/\omega)^k}$ 
        & $\exp\left[-(\nu/\omega)^k\right]$ 
        & $\omega,k > 0$ \\      
      \bottomrule
    \end{NiceTabular}
    }
    \label{sm:tab:distributions}
  \end{small}
\end{table*}

\clearpage
\section*{Supplementary references}

\end{document}